\documentclass{article}
\usepackage[utf8]{inputenc}
\usepackage[T1]{fontenc}
\usepackage{microtype}

\usepackage{amsmath,amsfonts,amssymb}
\usepackage{amsthm}
\usepackage{bm}
\usepackage{bbm}
\usepackage{nicefrac}
\usepackage{dsfont}

\usepackage{graphicx}
\usepackage{caption}
\usepackage{booktabs}
\usepackage{tabulary,multirow}
\usepackage{makecell}
\usepackage{tabularx}
\usepackage{tablefootnote}
\usepackage{array}
\usepackage{subcaption}

\usepackage{enumitem}
\usepackage{algorithm}
\usepackage{algorithmic}
\usepackage{thm-restate}

\usepackage[margin=1in]{geometry}
\usepackage{parskip}
\usepackage{authblk}
\usepackage{tcolorbox}

\usepackage[hyphens]{url}
\usepackage{hyperref}
\usepackage{cleveref}
\usepackage{natbib}

\usepackage{xcolor}

\definecolor{lightblue}{RGB}{173, 216, 230}

\newtcolorbox{definitioninstance}[1][Instance of Definition]{
  colback=lightblue!30,
  colframe=lightblue!70!black,
  boxrule=1pt,
  arc=4pt,
  title={#1},
  fonttitle=\bfseries
}

\theoremstyle{definition}

\title{How AI Impacts Skill Formation}
\author{Judy Hanwen Shen\thanks{Work done as a part of the Anthropic Fellows Program, \texttt{judy@anthropic.com}} \hspace*{1cm} 
Alex Tamkin\thanks{Anthropic, \texttt{atamkin@anthropic.com}}}
\date{\today}

\begin{document}

\maketitle
\begin{abstract}
AI assistance produces significant productivity gains across professional domains, particularly for novice workers. Yet how this assistance affects the development of skills required to effectively supervise AI remains unclear. Novice workers who rely heavily on AI to complete unfamiliar tasks may compromise their own skill acquisition in the process. We conduct randomized experiments to study how developers gained mastery of a new asynchronous programming library with and without the assistance of AI. We find that AI use impairs conceptual understanding, code reading, and debugging abilities, without delivering significant efficiency gains on average. Participants who fully delegated coding tasks showed some productivity improvements, but at the cost of learning the library. We identify six distinct AI interaction patterns, three of which involve cognitive engagement and preserve learning outcomes even when participants receive AI assistance. Our findings suggest that AI-enhanced productivity is not a shortcut to competence and AI assistance should be carefully adopted into workflows to preserve skill formation -- particularly in safety-critical domains. 
\end{abstract}
\section{Introduction}

Since the industrial revolution, skills in the labor market have continually shifted in response to the introduction of new technology; the role of workers often shifts from performing the task to supervising the task~\citep{autor2001skill}. For example, the automation of factory robots has enabled humans to move from manual labor to supervision, and accounting software has enabled professionals to move from performing raw calculations to identifying better bookkeeping and tax strategies. In both scenarios, humans are responsible for the quality of the final product and are liable for any errors~\citep{bleher2022diffused}. Even as automation changes the process of completing tasks, technical knowledge to identify and fix errors remains extremely important. 

As AI promises to be a catalyst for automation and productivity in a wide range of applications, from software engineering to entrepreneurship~\citep{dell2023navigating, peng2023impact,cui2024effects, otis2024uneven, brynjolfsson2025generative}, the impacts of AI on the labor force are not yet fully understood. Although more workers rely on AI to improve their productivity, it is unclear whether the use of AI assistance in the workplace might hinder core understanding of concepts or prevent the development of skills necessary to supervise automated tasks. Although most studies have focused on the end \emph{product} of AI assistance (e.g., lines of code written, quality of ideas proposed), an equally important, if not more crucial question is how \emph{process} of receiving AI assistance impacts workers. As humans rely on AI for skills such as brainstorming, writing, and general critical thinking, the development of these skills may be significantly altered depending on how AI assistance is used. 

Software engineering, in particular, has been identified as a profession in which AI tools can be readily applied and AI assistance significantly improves productivity in daily tasks~\citep{peng2023impact, cui2024effects}. Junior or novice workers, in particicular, benefit most from AI assistance when writing code. In high-stakes applications, AI written code may be debugged and tested by humans before a piece of software is ready for deployment. This additional verification that enhances safety is only possible when human engineers themselves have the skills to understand code and identify errors. As AI development progresses, the problem of supervising more and more capable AI systems becomes more difficult if humans have weaker abilities to understand code~\citep{bowman2022measuring}. When complex software tasks require human-AI collaboration, humans still need to understand the basic concepts of code development even if their software skills are complementary to the strengths of AI~\citep{wang2020human}. The combination of persistent competency requirements in high-stakes settings and demonstrated productivity gains from AI assistance makes software engineering an ideal testbed for studying how AI affects skill formation.

We investigate whether using and relying on AI affects the development of software engineering skills~\citep{handa2025economic}. Based on the rapid adoption of AI for software engineering, we are motivated by the scenario of engineers acquiring new skills on the job. Although the use of AI tools may improve productivity for these engineers, would they also inhibit skill formation? More specifically, does an AI-assisted task completion workflow prevent engineers from gaining in-depth knowledge about the tools used to complete these tasks? We run randomized experiments that measure skill formation by asking participants to complete coding tasks with a new library that they have not used before. This represents one way in which engineers acquire and learn new skills, since new libraries are frequently introduced in languages such as Python. We then evaluate their competency with the new library. Our main research questions are (1) whether AI improves productivity for a coding task requiring new concepts and skills, and (2) whether this use of AI reduces the level of understanding of these new concepts and skills. 
\begin{figure}
    \centering
    \includegraphics[width=\linewidth]{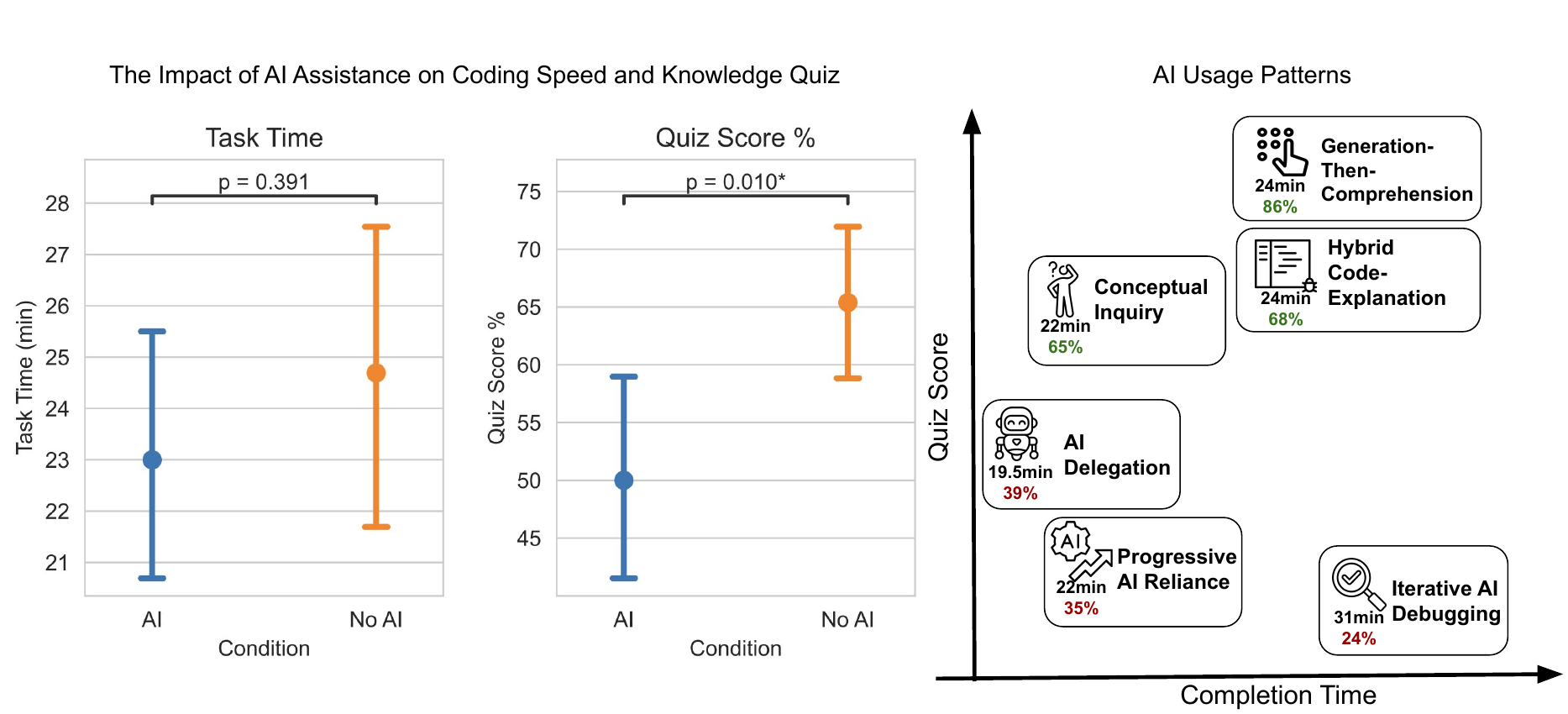}
    \caption{Overview of results: (Left) We find a significant decrease in library-specific skills (conceptual understanding, code reading, and debugging) among workers using AI assistance for completing tasks with a new python library. (Right) We categorize AI usage patterns and found three high skill development patterns where participants stay cognitively engaged when using AI assistance.}
    \label{fig:overview}
\end{figure}
\subsection{Our Results}
Motivated by the salient setting of AI and software skills, we design a coding task and evaluation around a relatively new asynchronous Python library and conduct randomized experiments to understand the impact of AI assistance on task completion time and skill development. We find that using AI assistance to complete tasks that involve this new library resulted in a reduction in the evaluation score by 15.4\% or two grade points (Cohen's $d=0.738$, $p=0.010$). Meanwhile, we did not find a statistically significant acceleration in completion time with AI assistance (Figure~\ref{fig:main-study}). 

Through an in-depth qualitative analysis where we watch the screen recordings of every participant in our main study, we explain the lack of AI productivity improvement through the additional time some participants invested in interacting with the AI assistant. Some participants asked up to 15 questions or spent more than $30\%$ of the total available task time on composing queries (Figure~\ref{fig:ai-interaction-time}). We attribute the gains in skill development of the control group to the process of encountering and subsequently resolving errors independently. We categorize AI interaction behavior into six common patterns and find three AI interaction patterns that best preserve skill development (Figure~\ref{fig:persona-overview}). These three patterns of interaction with AI, which resulted in higher scores in our skill evaluation, involve more cognitive effort and independent thinking (for example, asking for explanations or asking conceptual questions only).

\section{Background}
\subsection{The Impacts of AI Usage}

Since the widespread availability of ChatGPT, Copilot, Claude, and other advanced conversational assistants in late 2022, AI tools have been widely used in many different domains. Studies examining prompt-based utilization have facilitated a detailed examination of AI's real-world applications~\citep{tamkin2024clio, shen2025societal}. For example, AI tools are being used in professional domains such as software development, education, design, and the sciences~\citep{handa2025economic}. 

\paragraph{Productivity Gains} 
Many studies have found improvements in productivity using these AI assistants. For example, ~\citeauthor{brynjolfsson2025generative} found that AI-based conversational assistants increased the number of issues call center workers were able to resolve on average by 15\%. ~ \citeauthor{dell2023navigating} find similar results in which consultants completed 12.2\% more tasks on average with the help of AI than without it. While the skill-based effects differ across studies, a consistent pattern emerges in call center work, consulting, legal question-answering, and writing: less experienced and lower-skilled workers tend to benefit most~\citep{brynjolfsson2025generative, dell2023navigating, choi2023ai, noy2023experimental}. One exception was when GPT-4 was given to Kenyan small business owners, AI business advice helped high performers (by revenue) improve business results while worsening the results for lower performers~\citep{otis2024uneven}. 

For software engineering in particular, \citeauthor{peng2023impact} found that crowd-sourced software developers using copilot completed a task 55.5\% faster than the control group and novice programmers benefited more from AI coding assistance. Follow-up studies of developers in major software companies and found that AI-generated code completions provide a 26. 8\% boost in productivity as measured by pull requests, commits, and software product builds~\citep{cui2024effects}. This study also found that less experienced coders experienced greater boosts in productivity. While studies find that junior or less experienced developers experience greater productivity uplift from using AI, these very same workers should be quickly developing new skills in the workplace. Yet the effect of these tools on the skill formation of this subgroup remains unknown. Will the skill development of novice workers be affected significantly since they are still in the process of learning their trade? We are motivated by whether this productivity comes from free or at a cost.

\paragraph{Cognitive Offloading}
Concerns around the impact of AI assistance and skill depletion have been highlighted by recent works. For example, medical professionals trained with AI assistance might not develop keen visual skills to identify certain conditions~\citep{macnamara2024does}. In surveys given to knowledge workers, frequent use of AI has been associated with worse critical thinking abilities and increased cognitive offloading~\citep{gerlich2025ai}. Furthermore, knowledge workers reported a lower cognitive effort and confidence when using generative AI tools~\citep{lee2025impact}. However, these surveys are observational and may not capture the causal effects of AI usage. 

\paragraph{Skill Retention}
An adjacent line of inquiry to our research is how well humans retain knowledge and skills after AI assistance. ~\citeauthor{wu2025human} find that even when generative AI improved immediate performance on content creation tasks (e.g., writing a Facebook post, writing a performance review, drafting a welcoming email), the performance increase did not persist in subsequent tasks performed independently by humans afterward. For data science tasks, ~\citeauthor{wiles2024genai} described the impact of AI on non-technical consultants as an ``exoskeleton'', the enhanced technical abilities enabled by AI did not persist when workers no longer had access to AI. Our work asks the natural follow-up question of whether the usage of AI tools could cause worse learning outcomes for the acquisition of skills on the job for technical professionals themselves. 

\paragraph{Overreliance}
Although much of the literature in economics on AI-enhanced productivity implicitly assumes that generations of AI are trustworthy, the reality is that generative AI can produce incorrect~\citep{longwell2024performance} or hallucinated content~\citep{maleki2024ai}. When models are fallible, yet still deployed to assist humans, human decisions that follow erroneous model decisions are referred to as ``overreliance''~\citep{buccinca2021trust, vasconcelos2023explanations, klingbeil2024trust}. Although methods have been suggested to reduce overreliance, these focus mainly on decision-time information such as explanations~\citep{vasconcelos2023explanations, reingold2024dissenting} or debate~\citep{kenton2024scalable}. 

\subsection{CS Education and AI Assistance}
Measuring the acquisition of skills is highly domain dependent. For computer science in particular, most introductory courses measure learning through multiple choice questions, code writing, and code reading/explanations~\citep{cheng2022design}. More recent work has found code interviews, and active discussion of students' code to yield positive learning outcomes~\citep{kannam2025code}.  

Several observational studies have described how students use AI tools in the context of a computer science course. \citeauthor{poitras2024generative} found, over the course of a semester, that students used AI tools to write code, fix errors, and explain algorithmic concepts; students with less coding proficiency were more likely to seek AI assistance. Other works use surveys to find that students may be hesitant to use AI coding assistant tools due to ``dependence worry'' (i.e., overreliance on coding tools)~\citep{pan2024exploring}. For formal methods, \citeauthor{prasad2023generating} coded the different ways in which students used LLMs for course work and found that upper-year students taking the class did not rely on LLM assistance and only asked a few questions at the beginning. 

User studies have also been conducted in the professional development environments. \citeauthor{wang2024rocks} study different patterns in usage between users with and without chat access to AI models in completing coding puzzles and development tasks. They found rich interaction patterns including interactive debugging, code discussions, and asking specific questions. Participants ranged from asking ChatGPT to do then the entire problem (lowest quality code output) to only asking minimal questions (highest efficiency). Other studies have reported that AI tools help the software development process through easier access to documentation and accurate generation code for specific APIs~\citep{pinto2024developer}. 

\section{Framework}
\paragraph{Professional Skill Acquisition}
The ``learning by doing'' philosophy has been suggested by many learning frameworks such as the Kolb's experiential learning cycle, and the Problem-Based Learning (PBL)~\citep{kolb2014experiential, schmidt1994problem}. The frameworks connect the completion of real-world tasks with the learning of new concepts and the development of new skills. Experiential learning has also been explored specifically in software engineering courses in higher education in order to mimic solving problems in a professional setting~\citep{gonzalez2020experiential}. 
\begin{figure}
    \centering
    \includegraphics[width=0.7\linewidth]{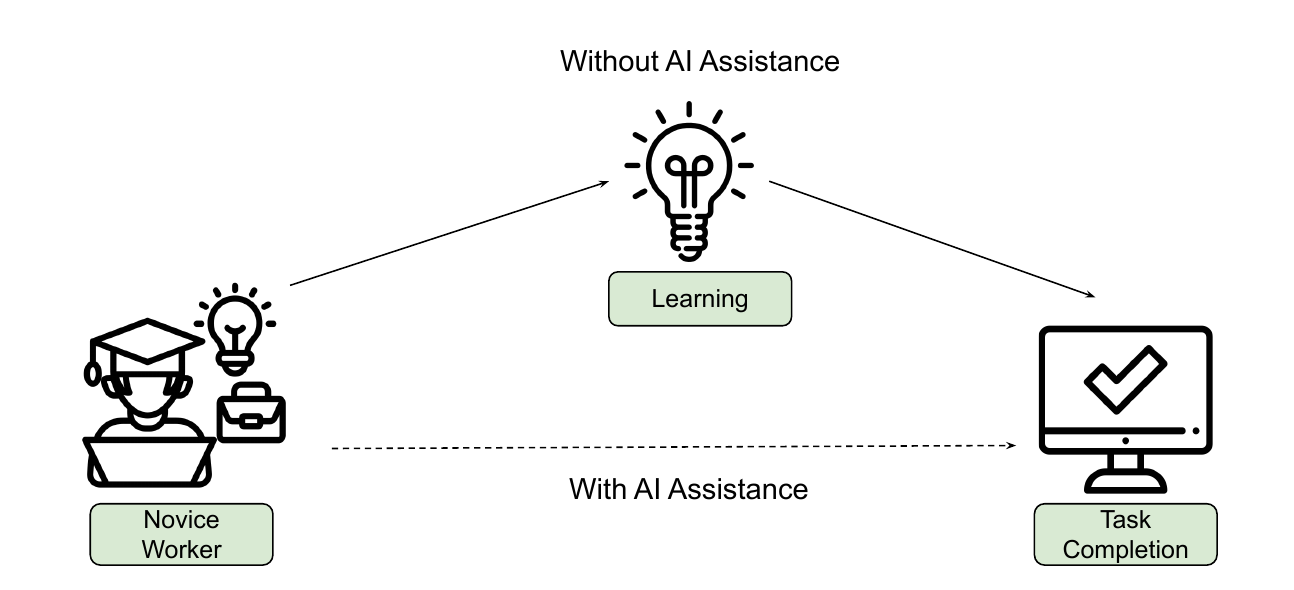}
    \caption{With AI assistance becoming more ubiquitous in the workplace, novice workers may complete tasks without the same learning outcomes. Our experiments aim to investigate the process of task completion requiring a new skill to understand the impact of AI assistance on coding skill formation.}
    \label{fig:framework}
\end{figure}
In its simplest form, we model AI tool assistance as taking a different learning path than without AI. We hypothesize that using AI tools to generate code in the development process effectively amounts to taking a shortcut to task completion without a pronounced learning stage. 

\paragraph{AI for Coding Usage Patterns} Prior works have found that humans use AI in many different ways for coding: from question answering to writing code, to debugging~\citep{poitras2024generative, wang2020human, pinto2024developer}. In our framework, different ways of using AI assistance represent different learning paths taken to reach the goals of completing the task. We analyze these different usage patterns in the qualitative analysis of this work (Section~\ref{sec:qual}). 

\paragraph{Research Questions}
Based on this background, we focus on on-the-job learning: settings where workers must acquire new skills to complete tasks. We seek to understand both the impact of AI on productivity and skill formation. We ask whether AI assistance presents a tradeoff between immediate productivity and longer-term skill development or if AI assistance presents a shortcut to enhance both. Our research questions are as follows:
\begin{itemize}
    \item \textbf{RQ1}: Does AI assistance improve task completion productivity when new skills are required? 
    \item \textbf{RQ2}: How does using AI assistance affect the development of these new skills? 
\end{itemize}

\section{Methods}
\subsection{Task Selection: Learning Asynchronous Programming with Trio}
We prototyped tasks for several different skills that junior software engineers may encounter on the job: from data analysis to plotting. We designed an experiment around the Python Trio library,\footnote{See documentation at: \url{https://trio.readthedocs.io/en/stable/}} which is designed for asynchronous concurrency and input-output processing (I/O). This library is less well known than \texttt{asyncio} (according to the number of \texttt{StackOverflow} questions) and involves new concepts (e.g., structured concurrency) beyond just Python fluency. It is also explicitly designed to be easy to use -- making it particularly suitable for a learning experiment.

\begin{figure}
    \centering
    \includegraphics[width=0.9\linewidth]{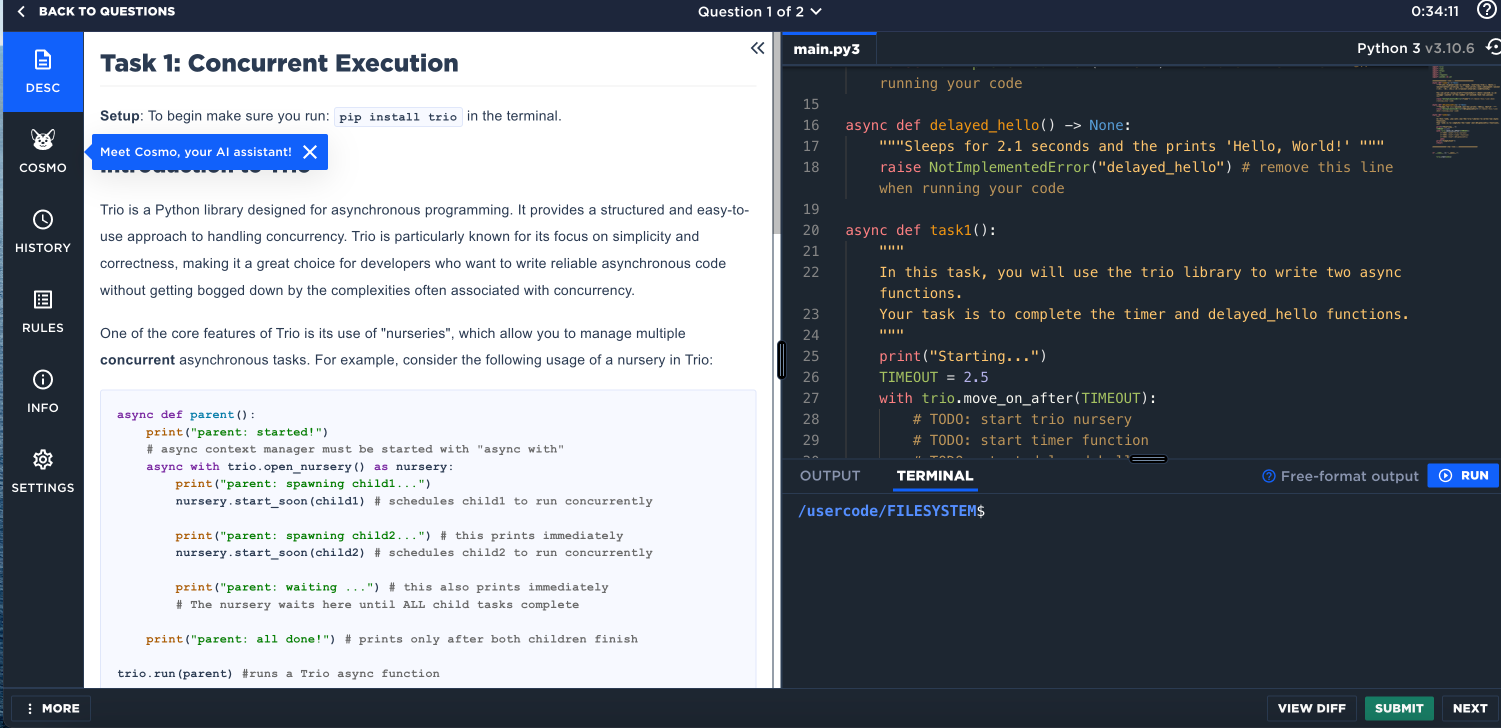}
    \caption{Experiment interface: We used a online interview platform to run our experiment. The treatment condition participants are prompted to use the AI assistant.}
    \label{fig:interface}
\end{figure}

We designed and tested five tasks that use the Trio library for asynchronous programming, a skill often learned in a professional setting when working with large-scale data or software systems. The tasks we created include problem descriptions, starter code, and brief descriptions of the Trio concepts required to complete the task. These tasks are designed to parallel the process of learning to use a new library or new software tool through a brief self-guided tutorial. For example, in software engineers' on-boarding materials, there is often a description of how to use an internal library and small tasks to build skills with the new library. 

After several pilot studies, we used the first two tasks in our main study; each task took 10 - 20 minutes during initial testing. The first task is to write a timer that prints every passing second while other functions run.  This task introduces the core concepts of nurseries, starting tasks, and running functions concurrently in Trio. The second task involves implementing a record retrieval function that can handle missing record errors in the Trio library. This task introduces concepts such as error handling and memory channels to store results. These two tasks are standalone; we provide sufficient instructions and usage examples so that participants can complete one task without the other. 

We used an online interview platform with an AI assistant chat interface (Figure~\ref{fig:interface}) for our experiments. Participants in the AI condition are prompted to use the AI assistant to help them complete the task. The base model used for this assistant is GPT-4o, and the model is prompted to be an intelligent coding assistant. The AI assistant has access to participants’ current version of the code and can produce the full, correct code for both tasks directly when prompted.

\subsection{Evaluation Design}
\label{sec:eval-design}
Based on a previous meta-analysis of evaluations in computer science education~\citep{cheng2022design}, we identify four types of questions used to assess the mastery of coding skills. Returning to our initial motivation of developing and retaining the skills required for supervising automation, proficiency in some of these areas may be more important than others for the oversight of AI-generated code. The four types of questions we consider are the following. 
\begin{itemize}
    \item \textbf{Debugging} The ability to identify and diagnose errors in code. This skill is crucial for detecting when AI-generated code is incorrect and understanding why it fails.
    \item \textbf{Code Reading} The ability to read and comprehend what code does. This skill enables humans to understand and verify AI-written code before deployment.
    \item \textbf{Code Writing} The ability to write or pick the right way to write code. Low-level code writing, like remembering the syntax of functions, will be less important with further integration of AI coding tools than high-level system design. 
    \item \textbf{Conceptual} The ability to understand the core principles behind tools and libraries. Conceptual understanding is critical to assess whether AI-generated code uses appropriate design patterns that adheres to how the library should be used.
\end{itemize} 
The two tasks in our study cover 7 core concepts from the Trio library. We designed a quiz with debugging, code reading, and conceptual questions that cover these 7 concepts. We exclude code writing questions to reduce the impact of syntax errors in our evaluation; these errors can be easily corrected with an AI query or web search. We tested 5 versions (Table 2) of the quiz in user testing and preliminary studies based on item response theory. For example, we ensure that all questions are sufficiently correlated with the overall quiz score, that each question has an appropriate average score, and that the questions are split up such that there is no local item dependence between questions (i.e., participants could not infer the answers to a question by looking at other questions). The final evaluation we used contained 14 questions for a total of 27 points. We submitted the grading rubric for the quiz in our study pre-registration before running the experiment. 
\subsection{Study Design}
\label{sec:study-design}

\begin{figure}
    \centering
    \includegraphics[width=0.8\linewidth]{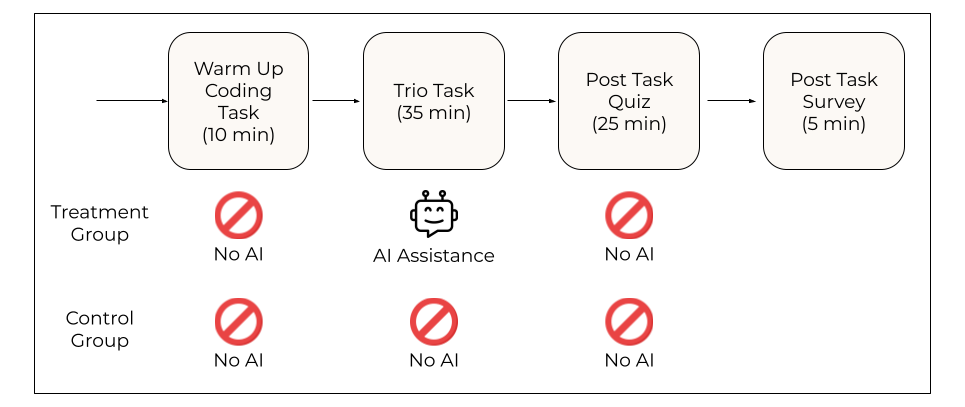}
    \caption{Overview of learning task and comprehension check. All participants completed a warm-up coding task that did not require Trio knowledge. During the main Trio task, participants in the treatment group could use AI assistance to answer questions or generate code. All participants were not allowed to use AI in the comprehension check.}
    \label{fig:taskflow}
\end{figure}
We use a between-subjects randomized experiment to test for the effects of using AI in the coding skill formation process. Each participant first completed a warm-up coding task on a coding platform, where they needed to add a border around a list of strings. This Python coding question takes an average of 4 minutes to complete among users of this coding platform. There are no asynchronous concepts in this coding question. No participants have access to AI while completing the warm-up stage. We use this stage to calibrate the Python familiarity of the participants and to help participants familiarize themselves with the interface.

The next stage is the Trio task stage, where participants have a maximum of 35 minutes to complete two coding tasks using Trio in the same coding platform. During this stage, participants in the AI assistance condition (treatment group) had access to coding help through a chat-based AI assistant (Figure~\ref{fig:interface}). All participants are instructed to complete the task as fast as they could. 

After completing the Trio task, participants completed the evaluation stage where they take the quiz we described in the previous section and complete a survey that consists of demographic and experiential questions after the quiz. 

In our main study, 52 participants completed the task, 26 for each of the control and treatment groups. For all our pilot studies and the main study, we only recruited participants who self-reported having more than one year of Python experience, code in Python at least once a week, have tried AI coding assistance at least a few times, and have never used the Trio library before (Table~\ref{tab:balancetable}). 
\begin{table}[]
    \centering
\begin{tabular}{lllll}
\toprule
            &                                  & Treatment (\%)       & Control (\%)        & Difference (\%) \\ \midrule
\multicolumn{2}{l}{Years of Coding Experience} &                         &                       &                   \\
            & 1-3 years                        & 2 (0.077)               & 2 (0.077)             & 0                 \\
            & 4-6 years                        & 10 (0.385)              & 9 (0.346)             & 1 (0.038)         \\
            & 7+ years                         & 14 (0.538)              & 15 (0.577)            & 1 (0.038)         \\
\multicolumn{2}{l}{Frequency of Python Use}    &                         &                       &                   \\
            & Regularly / Frequently           & 18 (0.692)              & 16 (0.615)            & 2 (0.077)         \\
            & Daily / Extensively              & 8 (0.308)               & 10 (0.385)            & 2 (0.077)         \\
\multicolumn{2}{l}{Per Task Async Quiz Score}  &                         &                       &                   \\
            & 0-2 (0-40\%)                     & 5 (0.192)               & 5 (0.192)             & 0                 \\
            & 3-4 (60-80\%)                    & 18 (0.692)              & 15 (0.577)            & 3 (0.115)         \\
            & 5 (100\%)                        & 3 (0.115)               & 6 (0.231)             & 3 (0.115)         \\
\multicolumn{2}{l}{Prior Python Asyncio Usage} & 18 (0.692)              & 20 (0.769)            & 2 (0.077)         \\
\multicolumn{2}{l}{Pre-Task Coding Time}       & 6.5 min & 8 min & 1.5 min   \\
\bottomrule
\end{tabular}
    \caption{Balance table of main study participants (n=52).}
    \label{tab:balancetable}
\end{table}

We use the coding platform to collect the keystrokes of the users as they code and the transcripts of their interaction with the AI coding assistant in the coding condition. We use Google Forms to collect survey responses from users both before the coding task and after the coding task. Together, these tasks take a maximum time of 1 hour and 15 minutes with an average duration of 58.5 minutes. Participants were recruited through a third party crowd-worker platform and paid a flat rate of 150 USD for the task.

\section{Results}
\subsection{Pilot Studies}

\begin{table}[]
    \centering
\begin{tabular}{llllp{9cm}}
\toprule
Pilot & Platform & Participants & No. Tasks & Challenges \\ \midrule
A     & P1    & n=39         & 5               & \textbf{Non-Compliance:} 35\% of participants in the no AI condition used AI assistance to copy the instructions and paste the results. Participants also self-reported using AI assistance in the no AI condition.                                                                                                                         \\
B     & P1    & n=107        & 5               & \textbf{Continued Non-Compliance:} Even when warned about the strict no AI requirements, participants continued to use AI for both coding and the quiz. 25\% of participants used AI and screen recording was not an option from the participant platform.                                                                                     \\
C     & P2   & n=20         & 5               & \textbf{Local Item Dependence:} Through watching screen recordings, we observed participants scrolling back and forth between questions to guess the correct answer.                                                                                                                                                                \\
D     & P2   & n=20         & 2               & \textbf{Python Syntax Delays:} In the time limit of 35 minutes, only 60\% of participants in the control (no AI condition) finished both tasks. The screen recordings showed several participants struggling with Python syntax issues, such as try/except blocks and string formatting. These delays were not germane to the Trio library. \\ \bottomrule
\end{tabular}
    \caption{Summary of pilot studies with different data providers, tasks, and evaluation design.}
    \label{tab:pilot}
\end{table}

\textbf{Non-Compliance} We conducted 4 pilot studies before running the full study (Table~\ref{tab:pilot}). The first two pilot studies were done on a different crowdworking platform (P1). On this platform, we observed a high level non-compliance (35\%) both during the task and the quiz (i.e., participants used AI to complete the coding task in the control group or used AI to complete the evaluation. We observed non-compliance behavior through the coding platform transcripts of when users copied the instructions or pasted code into the editor. We tested different mechanisms to ensure participants in the control condition (No AI) did not use AI for the task. However, despite more explicit instructions, around 25\% in the control group participants still used AI. We conducted two pilot studies with a second crowdworking platform (P2), each with 20 participants. Using screen recordings of participant progress, we verified that participants did not use AI in the control group nor for the quiz.

\textbf{Local Item Dependence} In Pilot Study C, we observed \textit{Local Item Dependence} in the quiz: participants would compare questions and identify answers based on code snippets provided in other questions. This motivated us to split the quiz into several different pages, where the questions on each page did not provide hints for other questions. Based on screen recordings, we observed that this reduced Local Item Dependence in pilot D. Additionally, we reduced the total number of tasks from five to two. This change allowed us to better isolate learning from the first two tasks while eliminating a confounding variable: participants in the AI condition seeing more concepts simply because they completed more tasks within the allotted time. To align the quiz with this modification, we adjusted the quiz questions to cover only the first two tasks.
\begin{figure}
    \centering
    \includegraphics[width=0.8\linewidth]{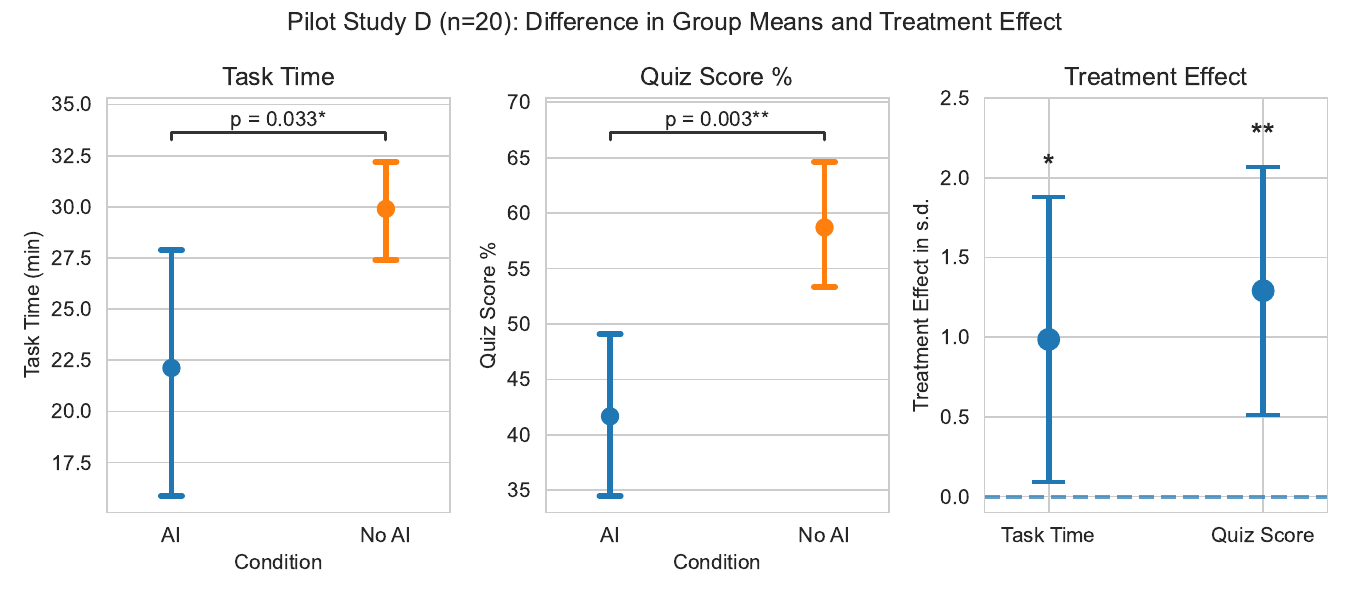}
    \caption{Difference in means of overall task time and quiz score between the control (No AI) and treatment (AI Assistant) groups in Pilot Study D. Error bars represent 95\% CI. Significance values correspond to treatment effect. * p<0.05, **<0.01, ***<0.001}
    \label{fig:pilot-results}
\end{figure}

\textbf{Barriers to Task Completion} In Pilot Study D, we included 20 participants. We found a significant difference in both the task completion time and the quiz score between the AI and non-AI conditions (Figure~\ref{fig:pilot-results}). When we reviewed the screen recording, participants in the control (no AI) condition struggled with Python syntax that was unrelated to Trio, such as \texttt{try/except} blocks and string formatting. The task competition rate within the 35-minute time limit was only 60\% within the control (no AI) group compared to a 90\% completion rate in the treatment (AI) group. Since our focus was not Python syntax, we added syntax hints about string formatting and 
\texttt{try/except} blocks for the main study. 

Figure~\ref{fig:pilot-results} presents the treatment effects on both outcome measures: Task Time and Quiz Score. The treatment (AI) group completed the Trio tasks faster (Cohen's d=1.11, p=0.03), demonstrating improved task efficiency. However, the AI group performed significantly worse on the knowledge quiz (Cohen's d=1.7, p=0.003), indicating reduced retention of learning. 
For our complete power analysis and pre-registration of the study,\footnote{Pre-registration: \url{https://osf.io/w49e7}} we assumed a conservative effect size of d = 0.85 (half of the observed learning effect) to account for the potential effect size inflation typical in pilot studies. 

\subsection{Main Study}
\subsubsection{Participants}
To recruit 50 participants, we sent our study to 58 crowd workers. Participants were balanced across the following attributes (recorded through a separate recruitment survey): years of coding experience, years of Python experience, prior usage of the Python Asyncio library, frequency of Python use in the past year, and an asynchronous programming familiarity score (a 5-question, multiple-choice concept check). The demographic breakdown of the participants, collected after the completion of the task to avoid the threat of stereotypes, is summarized in Figure \ref{fig:participant-distribution}. Most participants in our study hold a bachelor's degree, are between 25 and 35 years old, and work either as freelance or professional software developers. 53 participants completed all three parts of the study. Following our preregistered disqualification criteria, 1 participant was disqualified after leaving four blank questions on the quiz due to not realizing that there were multiple parts of the quiz and subsequently running out of time. 
\begin{figure}
    \centering
    \includegraphics[width=0.8\linewidth]{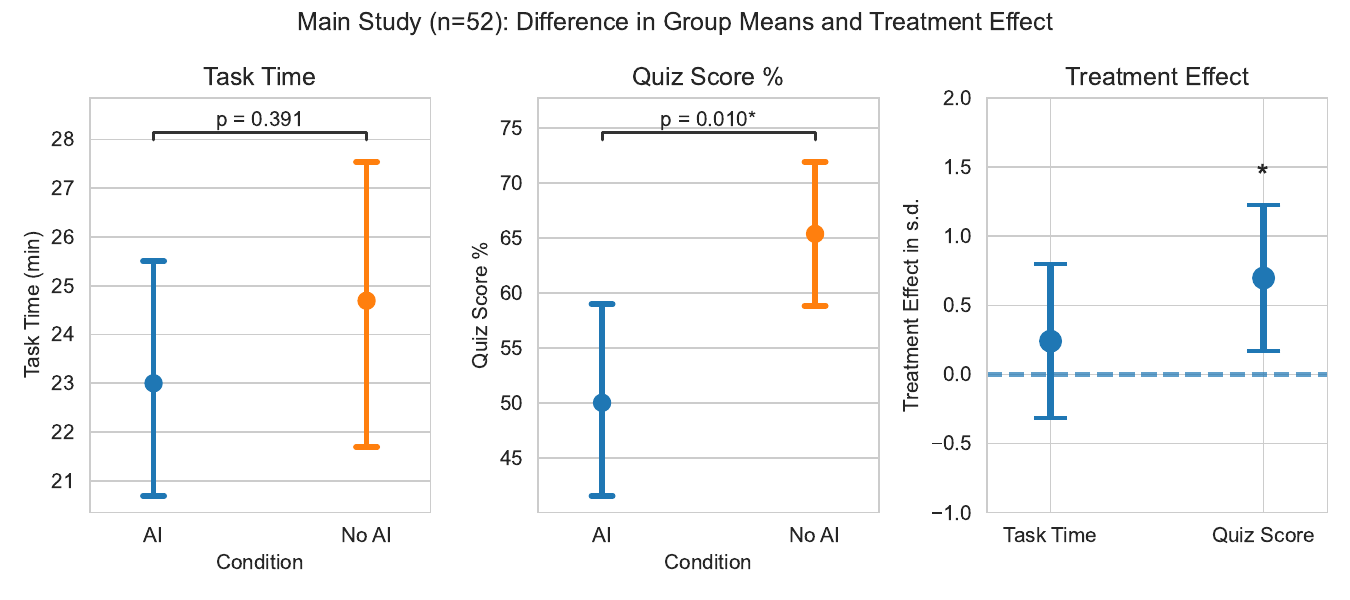}
    \caption{Difference in means of overall task time and quiz score between the control (No AI) and treatment (AI Assistant) groups in main study (n=52). Error bars represent 95\% CI. Significance values correspond to treatment effect. * p<0.05, **<0.01, ***<0.001
}
    \label{fig:main-study}
\end{figure}
\subsubsection{Results}
Figure~\ref{fig:main-study} shows that while using AI to complete our coding task did not significantly improve task completion time, the level of skill formation gained by completing the task, measured by our quiz, is significantly reduced (Cohen d=0.738, p=0.01). There is a 4.15 point difference between the means of the treatment and control groups. For a 27-point quiz, this translates into a 15.4\% score difference or 2 grade points.  Controlling for warm-up task time as a covariate, the treatment effect remains significant (Cohen’s d=0.725, p=0.016). 

\begin{figure}
    \centering
    \includegraphics[width=0.7\linewidth]{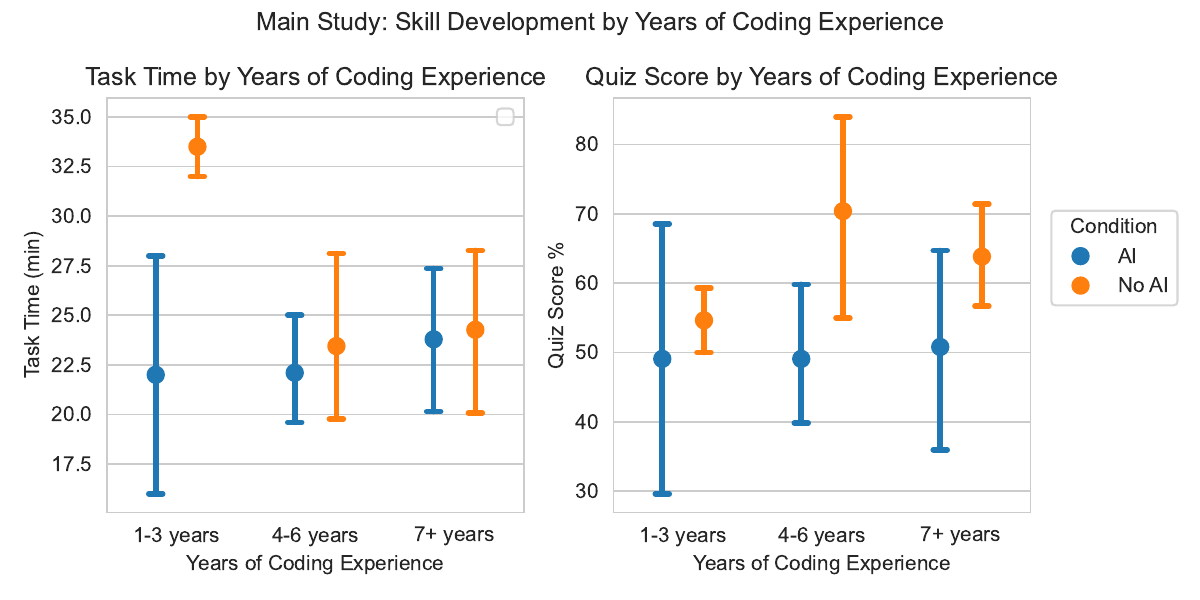}
    \caption{Task completion time and quiz score by years of coding experience. Error bars represent 95\% CI. The control group (No AI) average quiz score is higher across all levels of coding experience.}
    \label{fig:years_of_experience}
\end{figure}

Prior works have presented mixed results on whether AI helps or hinders coding productivity~\citep{peng2023impact, becker2025measuring}; our study differs from prior results in that it is designed to study how AI affects skill formation while performing a task requiring new knowledge. While we do observe a slightly lower average completion time in the AI group among novice programmers, due to the small group size of the 1-3 year participant group (n=4), the difference in task time was not significant. 4 of the 26 participants in the control (No AI) group did not complete the second task within the 35-minute limit, while every participant in the AI condition completed the second task. Our results do not conclusively find a speed up or slow down using AI in this task. 

Across all levels of prior coding experience, users scored higher on average in the control (no AI) than in the treatment (AI assistance) group (Figure~\ref{fig:years_of_experience}). This shows that our choice of tasks and task design did not critically hinge on the participants' experience level of the but presented new skills to be acquired for every experience group. 

\begin{figure}
    \centering
    \includegraphics[width=0.7\linewidth]{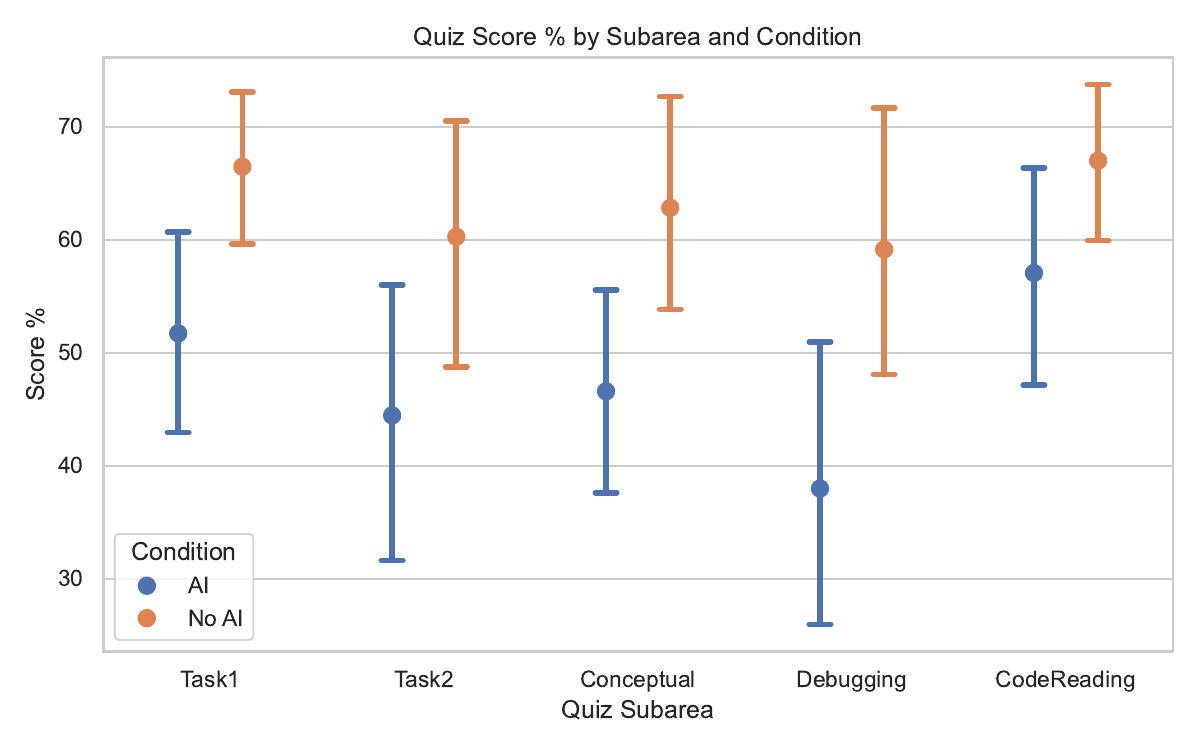}
    \caption{Score breakdown by questions type relating to each task and skill area. Debugging questions revealed the largest differences in average quiz score between the treatment and control groups.}
    \label{fig:quiz-score-subarea}
\end{figure}
\textbf{Concept Group Analysis} In exploratory data analysis (not pre-registered), the quiz score was decomposed into subareas and question types (Figure ~\ref{fig:quiz-score-subarea}). Each question in the quiz belonged to exactly one task (e.g., Task 1 or Task 2) and exactly one question type (e.g., Conceptual, Debugging, or Code Reading). For both tasks, there is a gap between the quiz scores between the treatment and control groups. Among the different types of questions, the largest score gap occurs in the debugging questions and the smallest score gap in the code reading questions. This outcome is expected since treatment and control groups may have similar exposure to reading code through the task, but the control group with no access to AI assistance encountered more errors during the task and became more capable at debugging.  


\begin{figure}[h]
  \centering
  \begin{minipage}{0.45\textwidth}
    \centering
    \includegraphics[width=\textwidth]{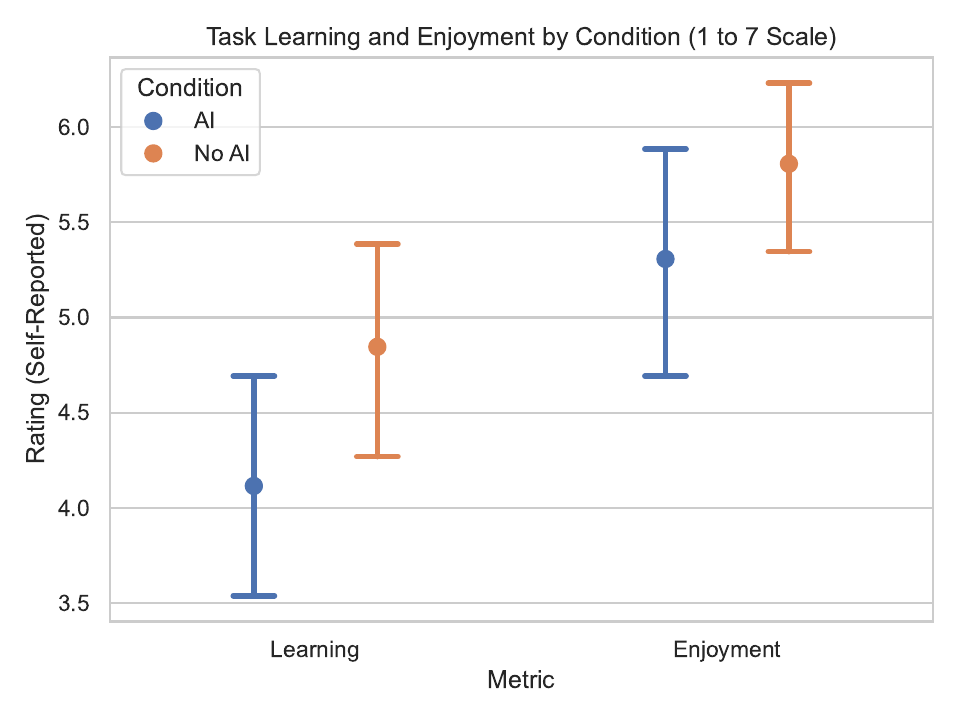}
    \caption{Self-reported enjoyment and learning by condition during our study. 
}
    \label{fig:enjoyment}
  \end{minipage}
  \hfill
  \begin{minipage}{0.45\textwidth}
    \centering
    \includegraphics[width=\textwidth]{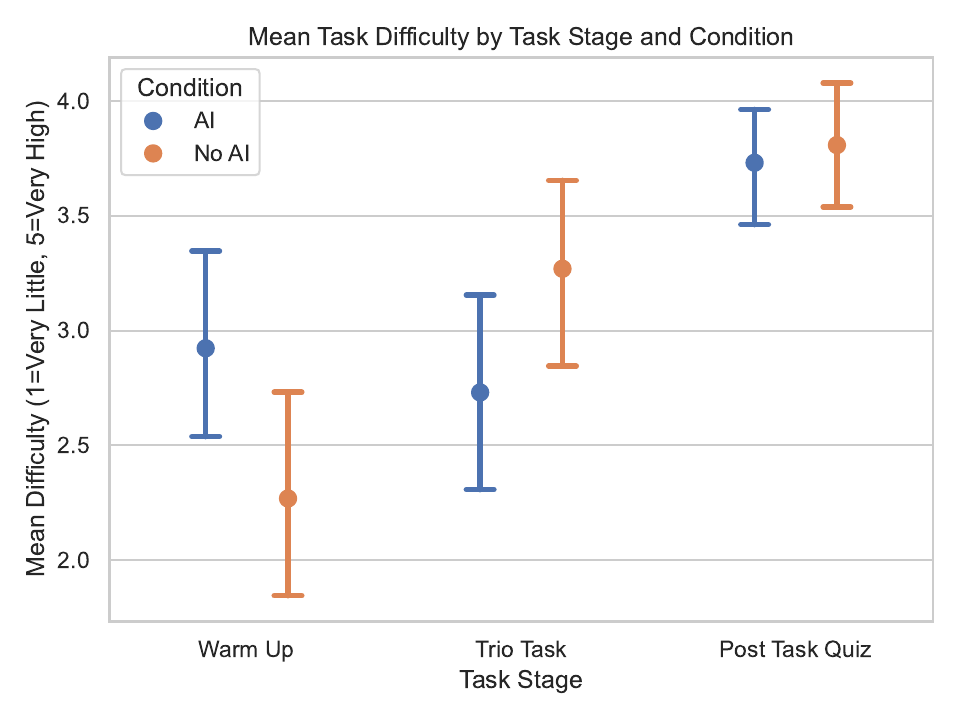}
    \caption{Self-reported task difficulty by condition during different stages of our study.}
    \label{fig:difficulty}
  \end{minipage}
\end{figure}

\textbf{Task Experience} In further exploratory data analysis, we also find differences in the way participants' experience of completing the study. The control group (No AI) reported higher self-reported learning (on a 7-point scale), while both groups reported high levels of enjoyment in completing the task (Figure~\ref{fig:enjoyment}). In terms of difficulty of the task, Figure~\ref{fig:difficulty} shows that although participants in the treatment group (AI Assistance) found the task easier than the control group, both groups found the post-task quiz similarly challenging.

\section{Qualitative Analysis}
\label{sec:qual}
\begin{figure}
    \centering
    \includegraphics[width=0.8\linewidth]{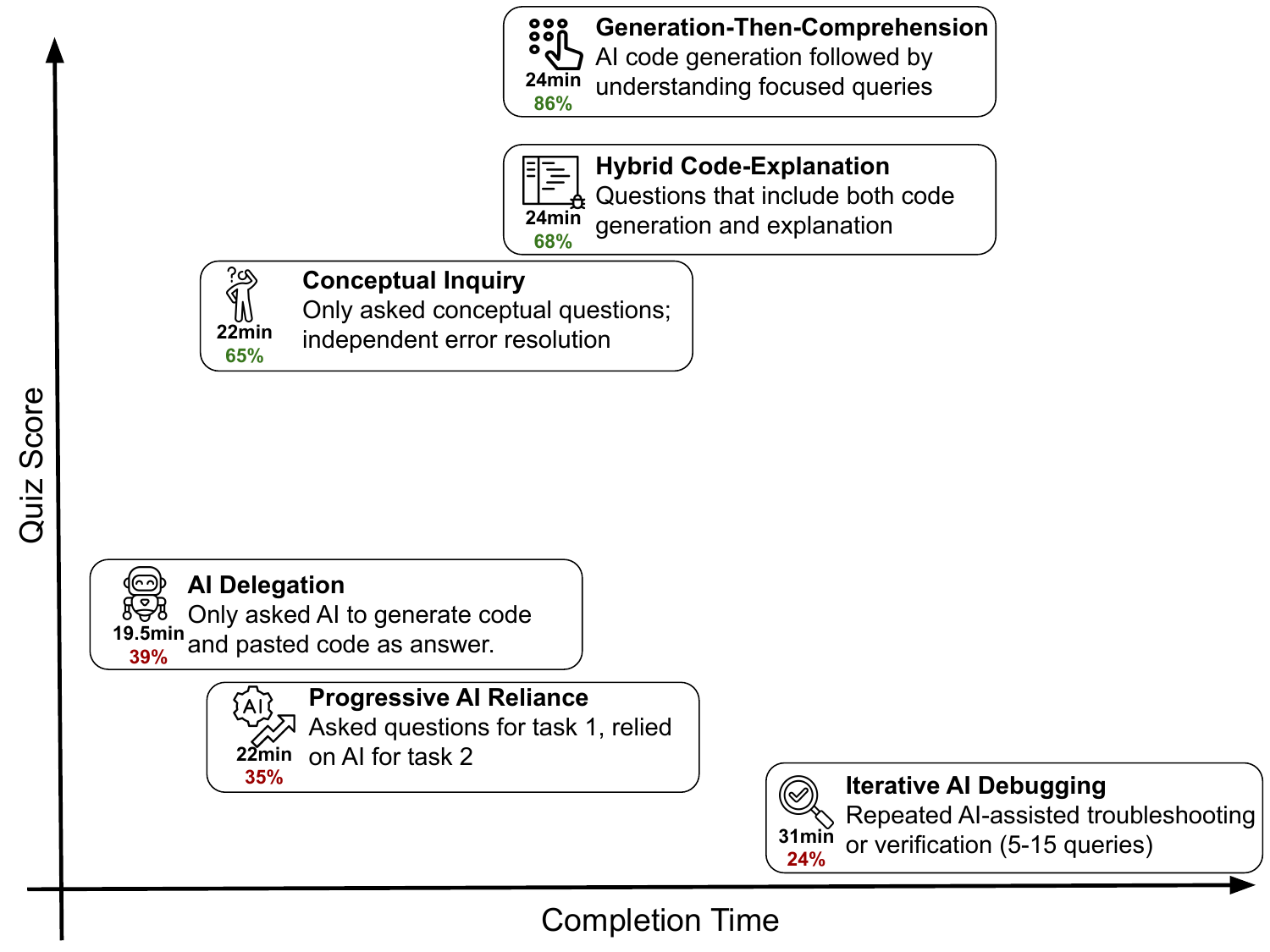}
    \caption{The 6 AI interaction personas in the treatment (AI) condition from our study with average completion times and quiz scores.}
    \label{fig:persona-overview}
\end{figure}

Although overall statistics on productivity and quiz score shed light on a high-level trend of how AI assistance affects a new learning task, a deeper analysis of how each participant completed the learning task allows us to better understand participant heterogeneity. In the initial coding phase of our qualitative analysis, we manually annotated screen recordings of the 51 participants in the main study.\footnote{The screen recording for one participant in the AI condition was not available.} We grouped the annotations into several main concepts related to task progress events such as errors, AI interactions, AI queries, and task completions (Table ~\ref{tab:app-annotation-guide}). This analysis allows us to understand not just the overall productivity and learning, but also how AI was used during each task in our study. We make these annotated transcripts publicly available for future studies.\footnote{Details on annotation procedures can be found in the Section~\ref{app:qual} and the annotated transcripts can be found at \url{https://github.com/safety-research/how-ai-impacts-skill-formation}} 

Analyzing these concepts or common patterns among participants helps supplement our quantitative observations of skill formation and task completion in this new library. Specifically, the following axes shows differences between participants and across conditions: 
\begin{itemize}
    \item \textbf{AI Interaction Time}: The lack of significant speed-up in the AI condition can be explained by how some participants used AI. Several participants spent substantial time interacting with the AI assistant, spending up to 11 minutes composing AI queries in total (Figure~\ref{fig:ai-interaction-time}). 
    \item \textbf{Query Types}: The study participants varied between conceptual questions only, code generation only, and a mixture of conceptual, debugging, and code generation queries. Participants who focused on asking the AI assistant debugging questions or confirming their answer spent more time on the task (Figure \ref{fig:ai_query_completion}). 
    \item \textbf{Encountering Errors}: Participants in the control group (no AI) encountered more errors;  these errors included both syntax errors and Trio errors (Figure~\ref{fig:error_type_all}). Encountering more errors and independently resolving errors likely improved the formation of Trio skills. 
    \item \textbf{Active Time}: Using AI decreased the amount of active coding time. Time spent coding shifted to time spent interacting with AI and understanding AI generations (Figure~\ref{fig:active-time}).
\end{itemize}

Using these axes, we develop a typology of six AI interaction patterns based on query types, number of queries, queries per task, and active time. As a result of this categorization, these six patterns yield different outcomes for both completion time and skill formation (i.e., quiz score). Figure~\ref{fig:persona-overview} summarizes each pattern and the average task outcomes. We can divide the interaction pattern into two categories: low- and high-scoring interaction patterns; the high-scoring patterns generally involve more cognitive effort and less AI reliance. Although each behavior pattern cluster is small, the difference between low-scoring clusters and high-scoring clusters is stark. One AI participant did not fit into any pattern clearly: they asked AI to implement everything for them in the first problem, then starts asking conceptual questions in the second task.   

\textbf{Low-Scoring Interaction Patterns} Low-scoring patterns generally involved a heavy reliance on AI, either through code generation or debugging. The average quiz scores in these groups are less than 40\%. Participants exhibiting these interaction patterns showed less independent thinking and more cognitive offloading~\citep{lee2025impact}. 
\begin{itemize}
    \item \textbf{AI Delegation} (n=4): Participants in this group wholly relied on AI to write code and complete the task. This group completed the task the fastest and encountered few or no errors in the process. 
    \item \textbf{Progressive AI Reliance} (n=4): Participants in this group started by asking 1 or 2 questions and eventually delegated all code writing to the AI assistant. This group scored poorly on the quiz largely due to not mastering any of the concepts in the second task. 
    \item \textbf{Iterative AI Debugging} (n=4): Participants in this group relied on AI to debug or verify their code. This group made a higher number of  queries to the AI assistant, but relied on the assistant to solve problems, rather than clarifying their own understanding. As a result, they scored poorly on the quiz and were relatively slower at completing the two tasks. 
\end{itemize}

\textbf{High-Scoring Interaction Patterns} High-scoring interaction patterns were clusters of behaviors where the average quiz score is 65\% or higher. Participants in these clusters used AI both for code generation, conceptual queries or a combination of the two. 
\begin{itemize}
    \item \textbf{Generation-Then-Comprehension} (n=2): Participants in this group first generated code and then manually copied or pasted the code into their work. After their code was generated, they then asked the AI assistant follow-up questions to improve understanding. These participants were not particularly fast when using AI, but demonstrated a high level of understanding on the quiz. Importantly, this approach looks nearly the same as the AI delegation group, but additionally uses AI to check their own understanding.  
    \item \textbf{Hybrid Code-Explanation} (n=3): Participants in this group composed hybrid queries in which they asked for code generation along with explanations of the generated code. Reading and understanding the explanations they asked for took more time.   
    \item \textbf{Conceptual Inquiry} (n=7): Participants in this group only asked conceptual questions and relied on their improved understanding to complete the task. Although this group encountered many errors, they also independently resolved these errors. On average, this mode was the fastest among high-scoring patterns and second fastest overall after the AI Delegation mode.   
\end{itemize}

\subsection{AI Interaction}
\label{sec:ai-interaction}
\paragraph{Interaction Time}
Contrary to previous work finding significant uplift or speedup of AI assistance for coding~\citep{peng2023impact, cui2024effects}, our results do not show a significant improvement in productivity if we \textit{only} look at the total completion time across the treatment and control groups. By analyzing how participants in the AI condition completed the task, the reason for the lack of improved productivity was due to the time spent interacting with the AI assistant. Some participants in the treatment group spent significant time (up to 11 minutes) interacting with the AI Assistant. For example, by typing or thinking about what to type. We capture the time invested in interacting with AI by labeling the time between when users start to type a query (annotated as an \textit{AI Interaction} event) and when an answer from the AI assistant is produced (annotated as an \textit{AI Query} event). 

\begin{figure}
    \centering
    \includegraphics[width=0.8\linewidth]{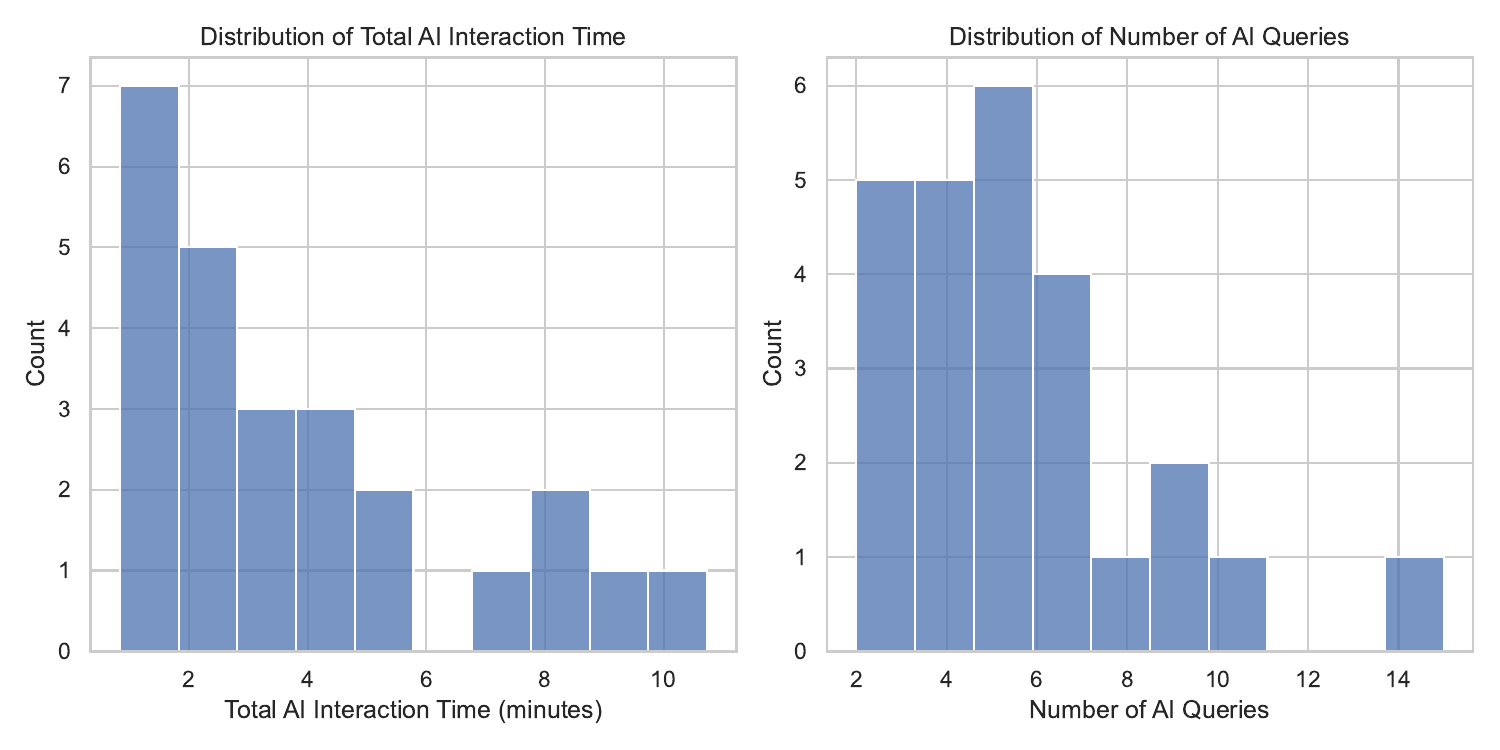}
    \caption{Distribution of total AI interaction time and number of AI queries. Participants spending more than 6 minutes interacting with AI during the task contribute to the treatment group (AI Assistance) not being significantly faster than the control group (No AI).}
    \label{fig:ai-interaction-time}
\end{figure}

Since participants could ask the AI assistant as many questions as time allowed, a handful of participants asked more than five questions and spent up to six minutes composing a single query during this 35-minute assignment (Figure~\ref{fig:ai-interaction-time}).\footnote{Participants were instructed to complete the task as fast as possible and were compensated a flat fee for participation. See Section~\ref{sec:study-design}.} Since the median completion time is only 19 minutes in the AI condition, spending up to 6 minutes composing a single query amounts to a significant amount of the total time spent interacting with the AI assistant. Although this effect might be due to the short duration of our task, \citeauthor{becker2025measuring} also found a slowdown effect for expert coders on longer tasks when participants waiting for AI-written code may become distracted. 

However, from the lens of skill formation, the time spent composing queries may aid in better understanding the task and, consequently, better acquisition of skills. Screen recordings show participants contemplating what to ask the AI assistant (e.g., rereading instructions and rewriting queries). As a result, some participants took several minutes to compose a single query. Thus, while this time cost would be more prominent in chat-based assistants than agentic coding assistants, the loss in knowledge is likely even greater in an agentic or autocomplete setting where composing queries is not required. A more significant difference in completion time due to shorter interactions with AI assistance would likely translate to an even larger negative impact on skill formation. When we look at individual queries, not all queries involve significant thinking and time. Thus, we analyze individual queries to better understand how participants from new skills.

\paragraph{AI Queries}
We categorized user inputs into the AI assistant, \textit{queries}, into 5 broad categories: explanation, generation, debugging, capabilities questions, and appreciation (Table \ref{tab:tab-queries}). The most common type of query was explanations (q=79); users requested more information about the trio library, details about asynchronous operations, and high-level conceptual introductions. 21 out of 25 participants in the treatment group asked an explanation question; this reflects the high level of engagement among our participants. The second most common were queries asking for code to be generated (q=51); some participants asked for an entire task to be completed, while other participants asked for specific functions to be implemented. Only 16 of 25 or two thirds of the participants used AI to generate code. 4 of these participants \emph{only} asked for code generation and no other types of question. In fact, 3 of the 8 lowest-scoring participants asked AI to generate code without asking for explanations, suggesting that if all participants in the AI group were to use AI for solely generating code, the skill-formation differences compared to the control group would be even greater.

A third category of common queries was debugging (q=9). Our tasks were designed to be straightforward, but the participants still encountered various errors (Section~\ref{sec:errors}). This is a broader category of queries that includes errors directly pasted as input to the AI assistant as well as asking the AI assistant to confirm the code written is correct. A higher fraction of debugging queries correlates with slower completion times (Figure \ref{fig:ai_query_completion}) and lower quiz scores (Figure \ref{fig:ai_quiz_score}). This suggests that relying on AI for debugging (e.g. repetatedly asking AI to check and fix things without understanding) when learning a new task is correlated with less learning.

Although we only recruited participants who have used AI assistants before, there were still questions (q=4) about whether the assistant could see the existing code and whether the assistant had knowledge of the specific library. In response to these questions, the AI assistance clarified that they could see the code and instructions. Several participants also expressed their appreciation for the assistant after the task was completed correctly. These expressions of appreciation, even at the cost of additional task time, reflect that politeness in the human-AI interaction~\citep{druga2017hey, ribino2023role} also appears in the context of AI for coding assistance. 

\setlength{\tabcolsep}{10pt} 
\renewcommand{\arraystretch}{1.5} 
\begin{table}[]
    \centering
    \begin{tabular}{p{3cm}|p{12cm}}
    \toprule
        Query Type & Example Query  \\ \midrule
        Explanation (q=79) & \textit{``can trio.sleep use partial seconds?''} \\ 
        & \textit{``Can you remind me what the different trio async operations are?''} \\ 
        & \textit{``Looks good, can you give me a really brief overview of the general idea behind all of this?''} \\ 
        \hline 
        Generation (q=51) & \textit{``given this instruction to trio, can you implement the missing bits of main.py?''}\\ 
        & \textit{``complete get\_user\_data''} \\ 
        & \textit{``implement delayed\_hello(). It should simply sleep for 2.1 seconds upon which it prints 'Hello World!' ''} \\ 
        \hline 
        Debugging (q=9) & \textit{``Does that look right? If so let's move on to delayed\_hello()''}\\ 
        & \textit{``I'm having issues getting my code to work. I'm getting a notimplementederror for delayed\_hello''} \\
        & Pasted Error (e.g., \textit{``Traceback (most recent call last):
        File "/usercode/FILESYSTEM/main.py3", line 81, in... ''}) \\ 
        \hline 
        Capabilities Question (q=4) & \textit{``Can you see the current question?''} \\ 
        & \textit{``So what can you do for me here? Can you write code directly into the file?''} \\
        & \textit{``Are you aware of how trio works? Are there parallels in its execution model to another library I'd be more familiar with like asyncio''} \\ \hline
        Appreciation (q=4) & \textit{``Great job, we got the expected output on the first try.''} \\ 
        & \textit{``Looks like it worked, thanks!''} \\ 
        & \textit{``Trueeee!''} \\
        \bottomrule
    \end{tabular}
    \caption{Examples of different types of queries received by AI assistant and counts of each type of query. 11 queries have multiple (two) labels.}
    \label{tab:tab-queries}
\end{table}

\begin{figure}
    \centering
    \includegraphics[width=0.7\linewidth]{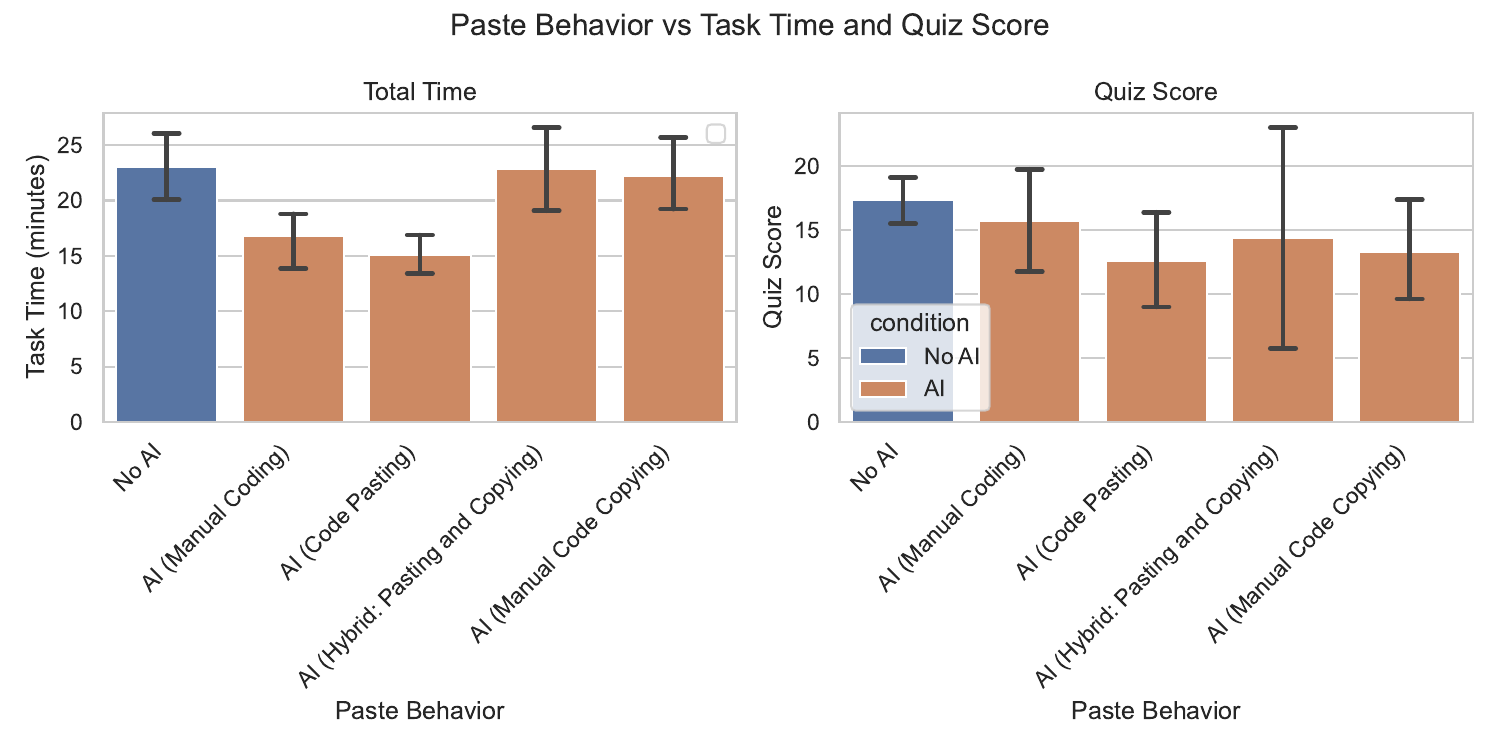}
    \caption{Decomposing AI Coding Behavior: Participants using AI by directly pasting outputs experience the most significant speed ups while participants who manually copied the AI-generated output were similar in pace to the control (No AI) group.}
    \label{fig:paste-behavior}
\end{figure}

\paragraph{Adopting AI Advice: Pasting vs Manual Code Copying} 
Another pattern that differs between participants is that some participants directly paste AI-written code, while other participants manually typed in (i.e., copied) the the AI generated code into their own file. The differences in this AI adoption style correlate with completion time. In Figure~\ref{fig:paste-behavior}, we isolate the task completion time and compare how the method of AI adoption affects task completion time and quiz score. Participants in the AI group who directly pasted ($n=9$) AI code finished the tasks the fastest while participants who manually copied ($n=9$) AI generated code or used a hybrid of both methods ($n=4$) finished the task at a speed similar to the control condition (No AI). There was a smaller group of participants in the AI condition who mostly wrote their own code without copying or pasting the generated code ($n=4$); these participants were relatively fast and demonstrated high proficiency by only asking AI assistant clarification questions. These results demonstrate that only a subset of AI-assisted interactions yielded productivity improvements. 

For skill formation, measured by quiz score, there was no notable difference between groups that typed vs directly pasted AI output. This suggests that spending more time manually typing may not yield better conceptual understanding. Cognitive effort may be more important than the raw time spent on completing the task. 

\subsection{Encountering Errors}
\label{sec:errors}
The way participants encountered and resolved errors was notably different between the treatment and control conditions. In the platform, participants could use the run button or the terminal to run their code as often as they wanted. In general, most of the participants ran the code for the first time after trying to complete most of the question and ran the code again only after the changes were made. We recorded every error encountered by each participant as we watched the screen recordings of the task progress. 

\begin{table}[]
    \centering
    \begin{tabular}{c|c|c} 
    \toprule
     & AI & No AI \\ \midrule
    Total &  1.0 (0.0-3.0) & 3.0 (2.0-5.0) \\
    Task 1 & 0.0 (0.0-2.0) & 2.0 (0.5-3.0) \\
    Task 2 & 0.0 (0.0-1.0) & 2.0 (0.5-2.0) \\ \bottomrule
    \end{tabular}
    \caption{Number of errors encountered per participant by condition. Values are median (Q1--Q3).}
    \label{tab:user-error}
\end{table}
\paragraph{Error Frequency} The AI group encountered fewer errors than the control group: the median participant in the treatment group encountered only one error in the entire task, while the median for the control group was three errors. Table \ref{tab:user-error} shows the difference in the error distributions. Most of the participants in the AI group were able to complete the tasks the first time they ran their code. In contrast, in the control condition, most of the participants encountered several errors in the process of completing each task. Among the 12 participants who completed both tasks without encountering errors, only two were in the control group. 

\paragraph{Errors and Trio Skill}
\begin{figure}[h]
  \centering
  \begin{minipage}{0.3\textwidth}
    \centering
    \includegraphics[width=\textwidth]{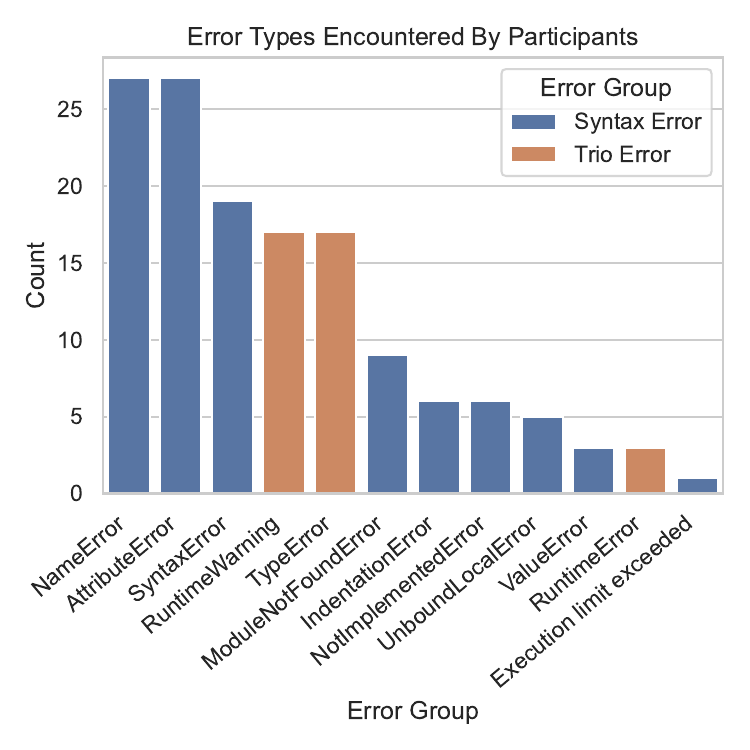}
    \caption{Count of all errors encountered by participants by error type.} 
    \label{fig:error_type_all}
  \end{minipage}
  \hfill
  \begin{minipage}{0.65\textwidth}
    \centering
    \includegraphics[width=\textwidth]{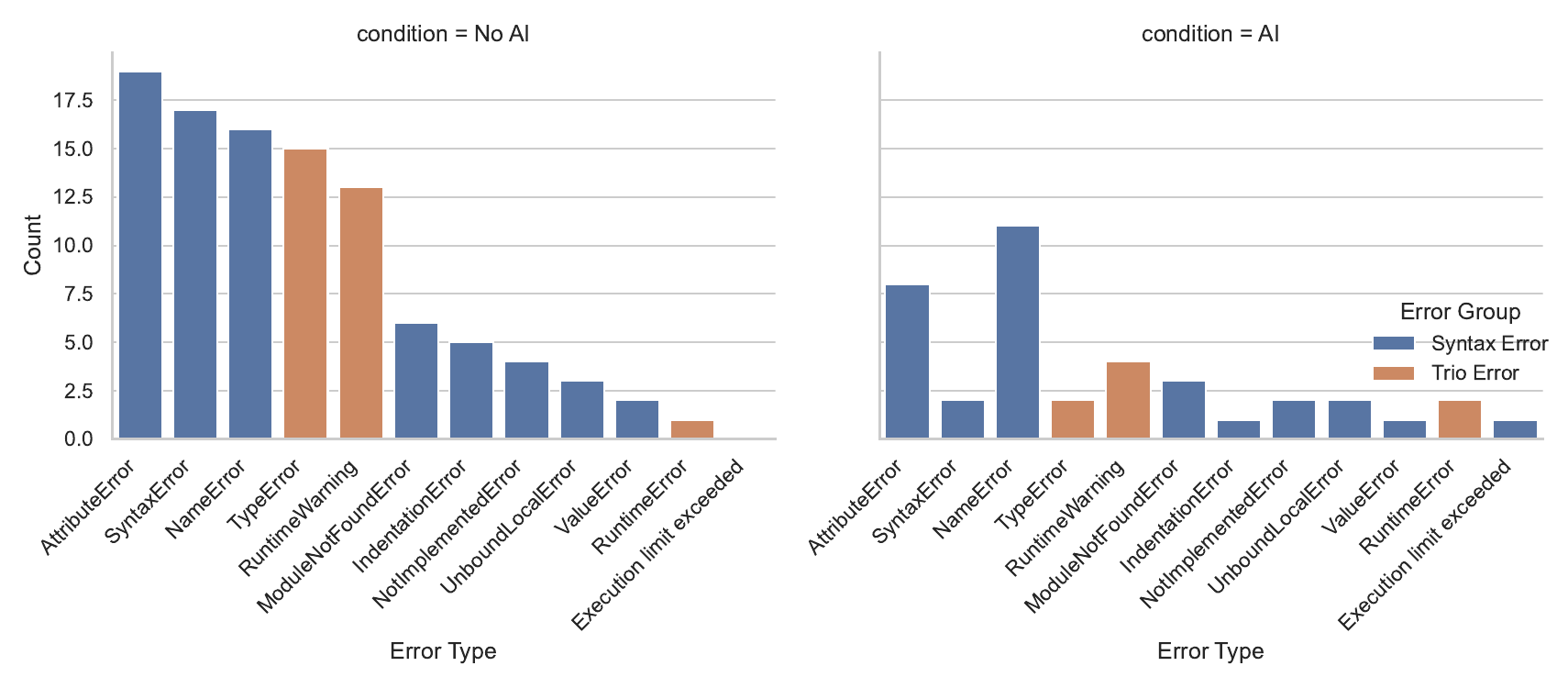}
    \caption{Count of errors by participant condition: AI (treatment) and No AI (control). The control group encountered many more errors related to key Trio concepts (e.g., \texttt{TypeError} and \texttt{RuntimeWarning}).}
    \label{fig:error_type_condition}
  \end{minipage}
\end{figure}

Not all errors carry the same weight in skill development in our study. Certain errors require a deeper understanding of the Trio library, which may account for differences in learning outcomes. Figure \ref{fig:error_type_all} shows that the most common errors are not directly related to the Trio library: \texttt{NameError} and  \texttt{AttributeError} are typically typos made on variable names or function names that are quickly corrected. Other errors are directly related to Trio: \texttt{RuntimeWarning} appears when a coroutine was never awaited and \texttt{TypeError} appears when a trio function gets a coroutine object instead of an async function. These errors force an understanding of key concepts on how the trio library handles corountines and the usage of await keywords that are tested in the evaluation. Although participants in the AI condition also encounter errors (Figure~\ref{fig:error_type_condition}), there are much fewer Trio-related errors encountered.

For participants in the control group, the higher frequency of encountering errors leads to more critical thinking about what is happening with the code and how to used the new library being presented. Furthermore, the frequent appearance of errors specifically related to the Trio library ensures that these specific concepts are gained in the process of completing the task. Together, these two differences suggest that encountering errors may play a significant role in the formation of coding skills. Moreover, our original motivation for the importance of preserving debugging skills may hinge on the acquisition of these skills without relying on AI. 

\subsection{Shifts in Active Coding Time}
\begin{figure}
    \centering
    \includegraphics[width=0.6\linewidth]{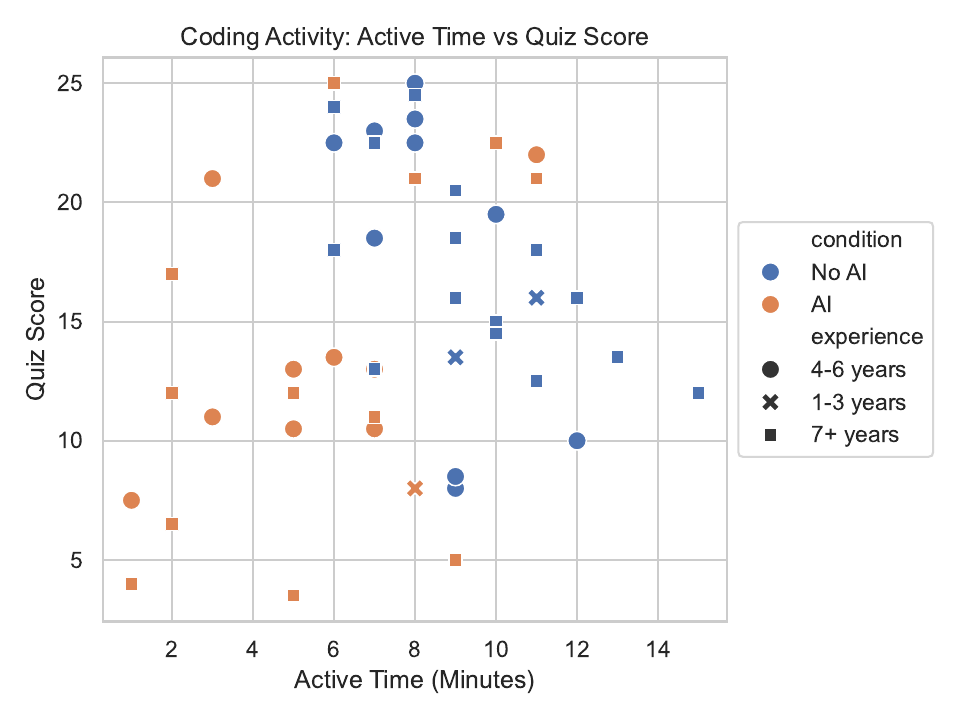}
    \caption{Active coding time vs. quiz score: active coding time represents the amount of time actually spent coding and is often a very small fraction of total task time. The No AI condition participants spent more active time coding and achieved higher quiz scores.}
    \label{fig:active-time}
\end{figure}

Although the outcome we measure in our main analysis is productivity through total task time, the actual amount to time spent actively coding illustrates a clearer pattern. Figure~\ref{fig:active-time} shows that participants in the AI condition spent much less active time on the task. This shift from coding to reading and understanding has also been found in previous work~\citep{becker2025measuring}. When we look at quiz score, the control group achieves high quiz scores with a higher total active time without the use of AI. Within each condition, higher active time correlates with lower quiz score, this is because the more experienced programmers spend less time actively coding while having better base knowledge compared to novice programmers. 

\subsection{Participant Feedback}
A quarter of the participants left feedback after the task and quiz were completed. In the control group (No AI), participants remarked that they found the task fun and that the tasks instructions were good at helping develop an understanding of Trio. In the treatment group (AI Assistance), participants remarked that they wished they had paid more attention to the details of the Trio library during the task, either by reading the generated code or by generating explanations in more depth. Specifically, participants reported feeling `lazy' and that `there are still a lot of gaps in (their) understanding'. The sentiment of participants' feedback suggested a more positive experience among the control group even though the task instructions and quiz questions were identical across groups (Table~\ref{tab:participant-feedback-AI} and Table~\ref{tab:participant-feedback-NoAI} provide all of the participant feedback from all participants). 

\section{Discussion}
Our main finding is that using AI to complete tasks that require a new skill (i.e., knowledge of a new Python library) reduces skill formation. In a randomized controlled trial, participants were assigned to the treatment condition (using an AI assistant, web search, and instructions) or the control condition (completing tasks with web search and instructions alone). The erosion of conceptual understanding, code reading, and debugging skills that we measured among participants using AI assistance suggests that workers acquiring new skills should be mindful of their reliance on AI during the learning process. Among participants who use AI, we find a stark divide in skill formation outcomes between high-scoring interaction patterns (65\%-86\% quiz score) vs low-scoring interaction patterns (24\%-39\% quiz score). The high scorers only asked AI conceptual questions instead of code generation or asked for explanations to accompany generated code; these usage patterns demonstrate a high level of cognitive engagement. 

Contrary to our initial hypothesis, we did not observe a significant performance boost in task completion in our main study. While using AI improved the average completion time of the task, the improvement in efficiency was not significant in our study, despite the AI Assistant being able to generate the complete code solution when prompted. Our qualitative analysis reveals that our finding is largely due to the heterogeneity in how participants decide to use AI during the task. There is a group of participants who relied on AI to generate all the code and never asked conceptual questions or for explanations. This group finished much faster than the control group (19.5 minutes vs 23 minutes), but this group only accounted for around 20\% of the participants in the treatment group. Other participants in the AI group who asked a large number of queries (e.g., 15 queries), spent a long time composing queries (e.g., 10 minutes), or asked for follow-up explanations, raised the average task completion time. These contrasting patterns of AI usage suggest that accomplishing a task with new knowledge or skills does not necessarily lead to the same productive gains as tasks that require only existing knowledge. 

Together, our results suggest that the aggressive incorporation of AI into the workplace can have negative impacts on the professional development workers if they do not remain cognitatively engaged. Given time constraints and organizational pressures, junior developers or other professionals may rely on AI to complete tasks as fast as possible at the cost of real skill development. Furthermore, we found that the biggest difference in test scores is between the debugging questions. This suggests that as companies transition to more AI code writing with human supervision, humans may not possess the necessary skills to validate and debug AI-written code if their skill formation was inhibited by using AI in the first place.


\subsection{Future Work} 
Our work is a first step to understanding the impact of AI assistance on humans in the human-AI collaboration process. We hope that this work will motivate future work that addresses the following limitations:
\begin{itemize}
    \item \textbf{Task Selection}: This study focuses on a single task using a chat-based interface. This should be a lower bound for cognitive offloading since agentic AI coding tools would require even less human participation. In our work, users who relied on AI without thinking performed the worst on the evaluation; a completely agentic tool would create a similar effect. Future work should investigate the impacts of agentic coding tools on learning outcomes and skill development. 
    \item \textbf{Task Length}: Ideally, skill formation takes place over months to years. We measured skill formation for a specific Python library over a one-hour period. Future work should study real-world skill development through longitudinal measurement of the impacts of AI adoption.
    \item \textbf{Participant Realism}: While participants in our study were professional or freelance programmers, there was not the same incentive to learn the library as if it were required for their actual job. Future studies should aim at studying the skill acquisition fro novice workers within a real company. 
    \item \textbf{Prompting Skills}: We collect self-reported familiarity with AI coding tools, but we do not actually measure differences in prompting techniques. An extension to our would also involve testing the level of prompting fluency beyond self-report. 
    \item \textbf{Evaluation Design}: Our study measures skill formation through a comprehensive quiz. Other studies could use the completion of another task or design coding as alternative evaluation strategies. 
    \item \textbf{Human Assistance}: We do not include the counterfactual of how skill formation would be impacted by receiving assistance from humans. Since human assistance and feedback takes place in a diverse settings (e.g., classroom, pair programming, code review), future work can compare the effect of feedback from AI vs humans in all these settings on skill formation. 
\end{itemize}

For novice workers in software engineering or any other industry, our study can be viewed as a small piece of evidence toward the value of intentional skill development despite the ubiquity of AI tools. Our study demonstrates the benefits of deploying cognitive effort when encountering a learning opportunity to master a new tool even if barriers (e.g., errors) may be encountered in the process of mastery. Exerting cognitive effort can be assisted by AI; beyond the patterns we describe, major LLM services also provide learning modes (e.g., ChatGPT Study Mode,  Claude Code Learning / Explanatory mode). Ultimately, to accommodate skill development in the presence of AI, there needs to be a more expansive view of the impacts of AI on workers. Participants in the new AI economy must care not only about productivity gains from AI but also the long-term sustainability of expertise development amid the proliferation of new AI tools. 

\section{Acknowledgments}
We would like to thank Ethan Perez, Miranda Zhang, and Henry Sleight for making this project possible through the Anthropic Safety Fellows Program. We would also like to thank Matthew J\"orke, Juliette Woodrow, Sarah Wu, Elizabeth Childs, Roshni Sahoo, Nate Rush, Julian Michael, and Rose Wang for experimental design feedback. 

We like to thank Jeffrey Shen, Aram Ebtekar, Minh Le, Mateusz Piotrowski, Nate Rahn, Miles McCain, Jessica Zhu, Alex Wang, John Hewitt, Rosanne Hu, Saffron Huang, Kyle Hsu, Sanjana Srivastava and Jennifer Leung for task testing and feedback. 

\newpage
\bibliography{references}
\newpage
\appendix
\section{Participant Details}
\begin{figure}
    \centering
    \includegraphics[width=\linewidth]{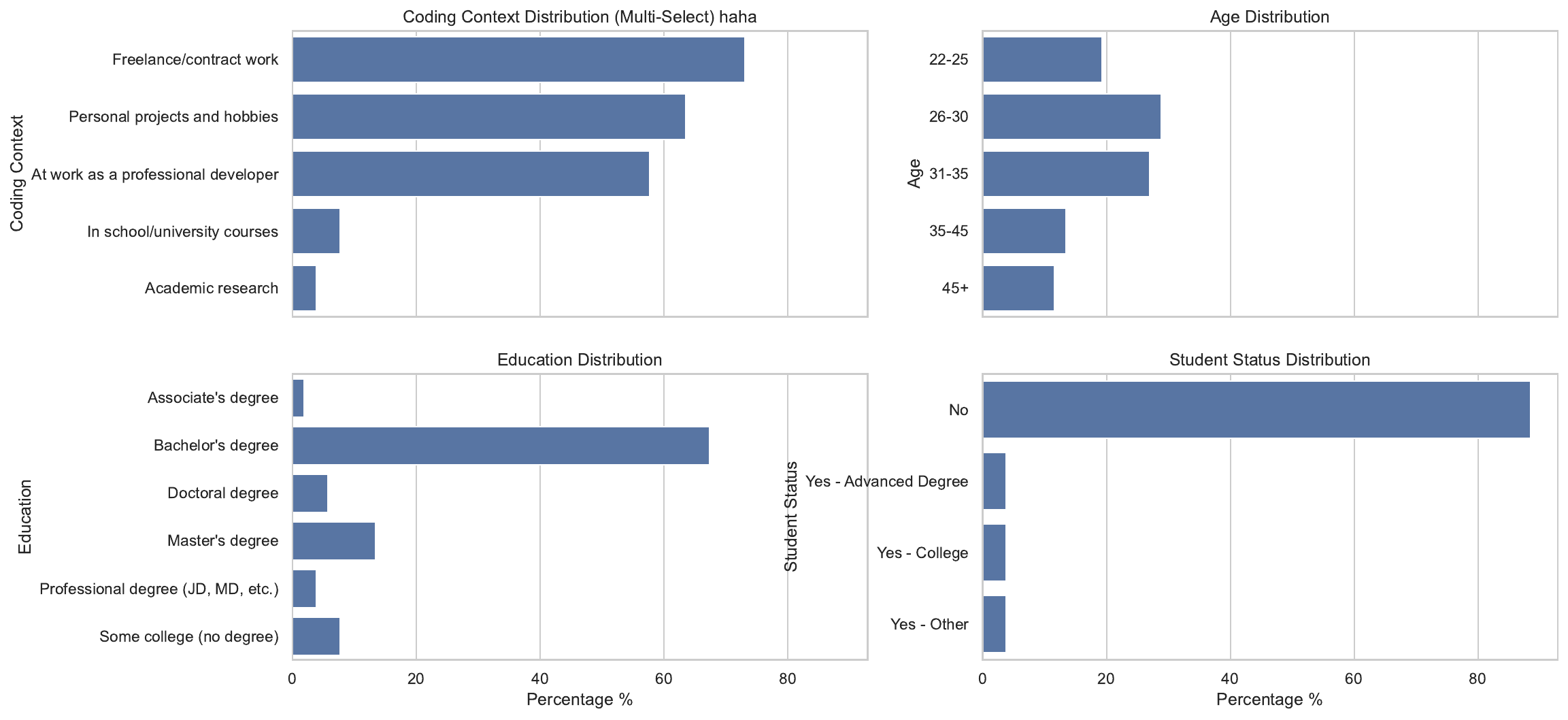}
    \caption{Participant distribution for main study, collected after the task to avoid stereotype threat. Most participants are professional programmers.}
    \label{fig:participant-distribution}
\end{figure}

\subsection{Ethics Review}
The protocol was reviewed and approved by internal reviewers at Anthropic. Participants were not exposed to any risks during this study. The benefits which may reasonably be expected to result from this study are learning a new software (Python) skill. We did not guarantee or promise that participants will receive any specific learning benefits from this study. We collected informed consent for participation in the study during the prescreening stage. We gave participants the right to withdraw consent at any time without penalty. They will still be compensated even if they fail the attention checks, do not fully complete the task, or complete the task incorrectly. 
Quiz responses are stored in Google Drive, and coding keystrokes are stored in the coding Platform. All stored information is completely anonymized; only the data platform can use the IDs to identify the participants for payment. We further remove data platform identifiers to annotate coding patterns between the two groups.

\section{Qualitative Analysis Data and Details}
\label{app:qual}
\subsection{Annotation Procedure}
51/52 participants uploaded screen recordings of their work in the warm-up task, main coding task, and quiz. We watched the recordings of for all participants (25 AI condition, 26 no AI condition) for the main coding task. We record the time stamps of the following events: 
\begin{table}[]
    \centering
    \begin{tabular}{l|l|l}
    \toprule
      Event    &  Description & Additional Info\\
      \midrule
      Task Start  & User opens each task & \\ 
      AI Interaction & User starts typing into AI window & Description of interaction \\ 
      AI Query & User receives answer from AI & Query\\ 
      Websearch & User queries search engine & Query\\
      Paste (Direct) & User pastes output of AI assistant & \\
      Code Copying & User types code using AI output & \\
      Error & Code produces error when run & Error Message\\
      Interface Error & Development environment or AI assistant error & \\ 
      Task Completion & Correct output is achieved & \\
      Task Submission & User submits task & Code completion \\
      \bottomrule
    \end{tabular}
    \caption{Events annotated manually for each video recording of the main task.}
    \label{tab:app-annotation-guide}
\end{table}

We also note general themes in how participants use AI in each condition based on these codes of events in the event.

\begin{figure}[h]
  \centering
  \begin{minipage}{0.8\textwidth}  
    \centering
    \includegraphics[width=\textwidth]
    {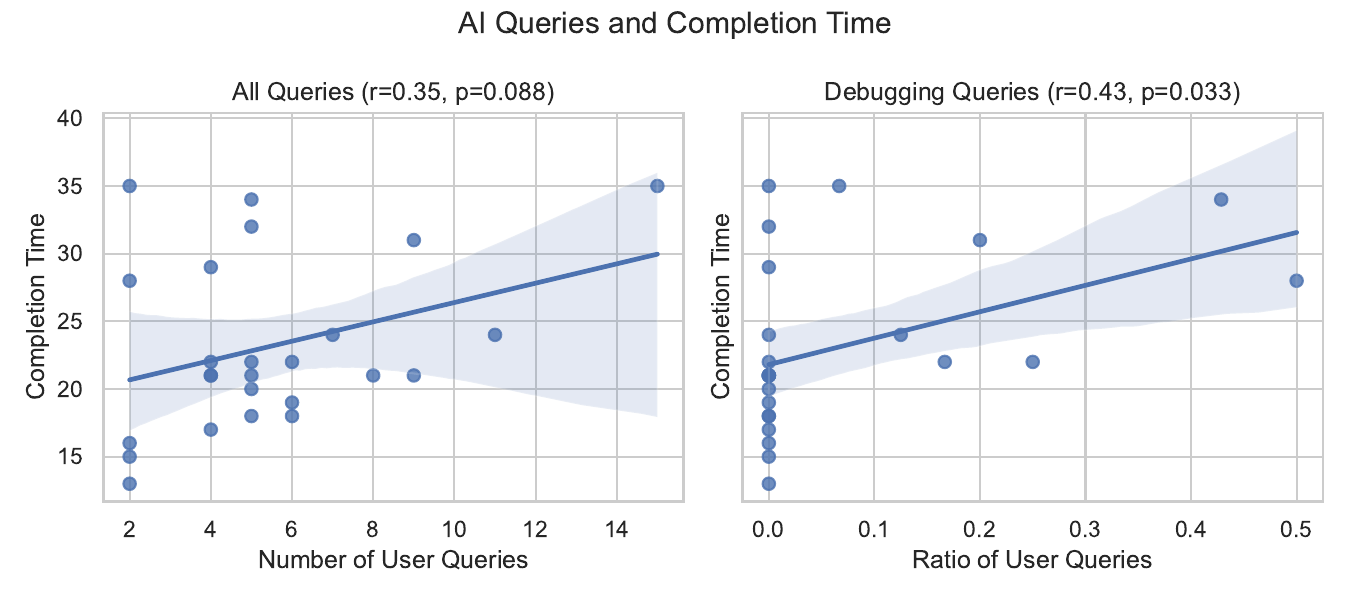}
    \caption{As the number of queries the total completion time increases,  users who ask a high fraction of debugging queries also tend to use more time to complete the task.}
    \label{fig:ai_query_completion}
  \end{minipage}
  
  \vspace{1em}  
  
  \begin{minipage}{0.8\textwidth}
    \centering
    \includegraphics[width=\textwidth]{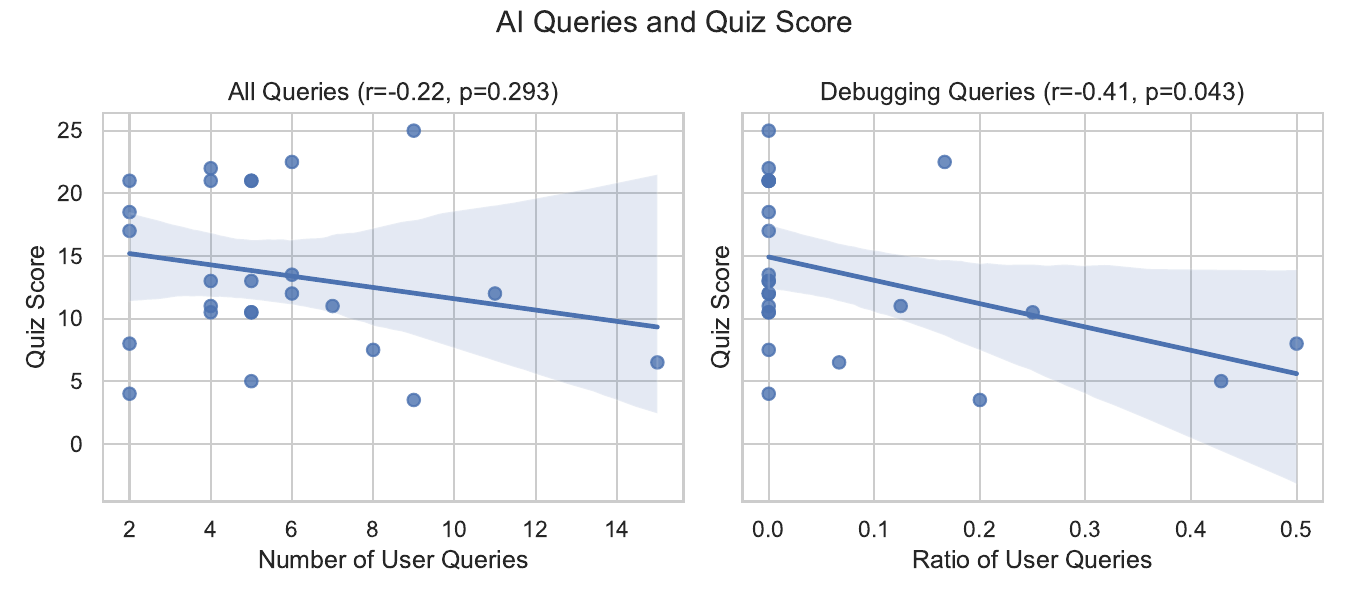}
    \caption{There is no clear pattern between the number of total queries and quiz score. However, users who heavily rely on AI to debug tend to have lower quiz scores.}
    \label{fig:ai_quiz_score}
  \end{minipage}
\end{figure}

\subsection{Data Availability}
We make the annotated transcripts of each participant available at the following URL: \url{https://github.com/safety-research/how-ai-impacts-skill-formation}. 

\subsection{Participant Feedback Details}
We include all participant feedback in Table~\ref{tab:participant-feedback-AI}
 and \ref{tab:participant-feedback-NoAI}. Since the average completion time was faster for the AI group, the AI group left more comments since they felt like they had more time at the end of the task. 
 \begin{table}[]
    \centering
    \begin{tabular}{c|p{11cm}}
    \toprule
    Condition & Feedback \\ \midrule
     AI    &  The first quiz (printing the *s) would have been much easier if I could just look up the syntax of appending to an array. I forgot it when doing the task - and normally, that'd take five seconds to google and would require no model assistance, but since I couldn't leave the tab, I got stuck. \\
     AI     &  By using the AI assistant, I feel like I got lazy. I didn't read the Trio library intro and code examples as closely as I would have otherwise. \\
     AI & Very cool project, hope it goes well! \\ 
     AI & I would not have minded going into a little more depth with the assistant to really understand and prove out the details of the trio library. I feel like I got a pretty good overview but there are still a lot of gaps in my understanding. \\
     AI & I feel stupid from the warmup, but hopefully the other project demonstrated what I can do. Sorry if I wasted your time. \\
     AI & It was fun and challenging. The warm up was confusing because it seemed the task had some issues but overall the entire thing was a fun learning experience. \\ 
     AI & I wish I'd taken the time to understand the explanations from Cosmo a bit more! \\
     AI & I'm slow. I think the time limit made me act in a way that wasn't representative to my normal workflow, particularly in the proportion of time spent building mental models vs. obtaining code progressions. I had the desire to understand trio a lot more than I allowed myself in the moment, because I knew I wouldn't have the time. I also think my attention was scattered as I tried to operate in a rush. Trio seems to work pretty similarly to other libraries I am realizing I'm not as familiar with as I thought. My sense after this is that there maybe be differences in the execution model, but I didn't really get to dig deep enough to understand them. I'm surprised that something as simple as understanding the `start\_soon` method... like I picked up nothing about that in terms of deeper understanding. Thanks for letting me participate! \\ 
     AI & this was fun! I wish i paid more attention to trio syntax when coding with Cosmo \\ 
     \bottomrule
    \end{tabular}
    \caption{Feedback from Participants in the AI Condition}
    \label{tab:participant-feedback-AI}
\end{table}

\begin{table}[]
    \centering
    \begin{tabular}{c|p{11cm}}
    \toprule
    Condition & Feedback \\ \midrule
     No AI    & This was a lot of fun but the recording aspect can be cumbersome on some systems and cause a little bit of anxiety especially when you can't go back if you messed up the recording. \\
     No AI     & It was fun to learn about asynchronous programming, which I had not encountered before. I think I could have done much better if I could have accessed the coding tasks I did at part 2 during the quiz for reference, but I still tried my best. I ran out of time as the bug-finding questions were quite challenging for me.  \\
     No AI     &  This was fun \\
No AI     &  The programming tasks were very fun and did a good job of helping me understand how Trio works despite never having used it before. I spent too much time on this quiz, but that was due to my time management. Even if I hadn't spent too much time on the first part, though, it still would have been a tight finish for me in the 30 minute window I think. \\
     \bottomrule
    \end{tabular}
    \caption{Feedback from Participants in the No AI (control) Condition}
    \label{tab:participant-feedback-NoAI}
\end{table}

\section{Evaluation Details}
\label{app:eval-details}

\subsection{Evaluation Design}
\begin{figure}
    \centering
    \includegraphics[width=\linewidth]{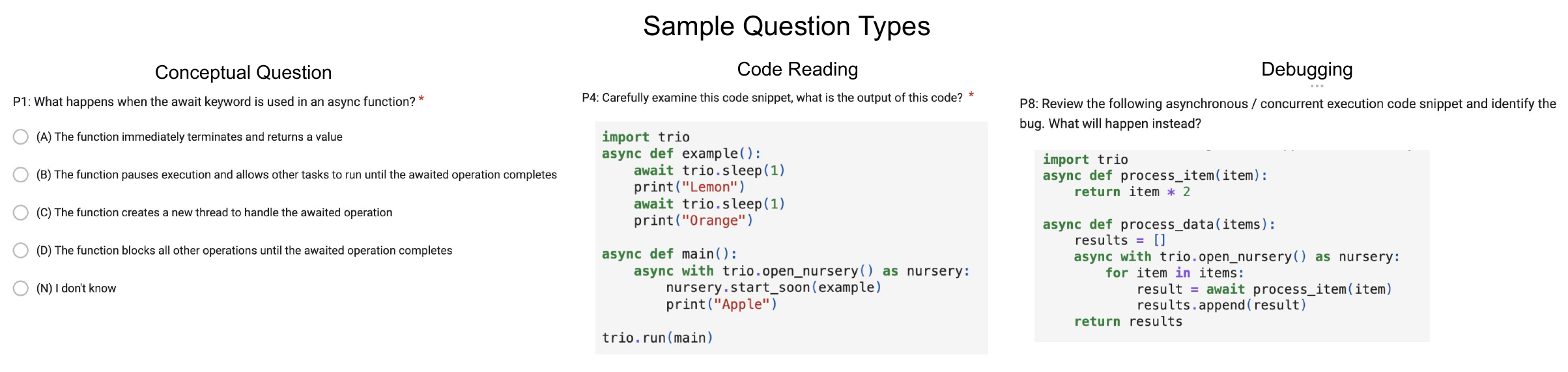}
    \caption{Example question types from our evaluation. We designed the evaluation to test three different software skills: conceputal understanding, code reading, and code writing.}
    \label{fig:question-types}
\end{figure}
\textbf{Question Types} We discuss the three types of questions we used: Conceptual Understanding, Code Reading, and Debugging in Section~\ref{sec:eval-design}. 

\textbf{Knowledge Categories}
The evaluation covers 7 core concepts from the Trio library: 
\begin{enumerate}
    \item \textbf{Async and await keywords}: When to use await keywords within async functions. For example: ``What happens when the await keyword is used in an async function?''
    \item \textbf{Starting Trio functions}: Basic Trio usage including how to spawn tasks and how spawned tasks with different durations behave. 
    \item \textbf{Error handling in Trio}: Understanding error propagation patterns and how to catch errors in child tasks. For example `` What happens to a parent task when a child task raises an unhandled exception in Trio?'' 
    \item \textbf{Coroutines:} When calling async functions, how to debug co-routine never awaited errors. 
    \item \textbf{Memory channels using Trio}: Understanding that \texttt{start\_soon} doesn't return anything and how to use dictionaries, lists and other memory channels to collect data when running multiple tasks in parallel.  
    \item \textbf{Opening and closing a Trio nursery}: Understanding asynchronous context managers and how to use them. For example, a debugging question consisting of a snippet of code where the nursery is started in correctly. 
    \item \textbf{Sequential vs concurrent execution:} The expected behavior of concurrent tasks. For example ``Read the following code and identify when each task starts and completes''
\end{enumerate}

\section{Task Details}
\label{app:task-details}

\begin{figure}[h]
  \centering
  \begin{minipage}{0.45\textwidth}
    \centering
    \includegraphics[width=\textwidth]{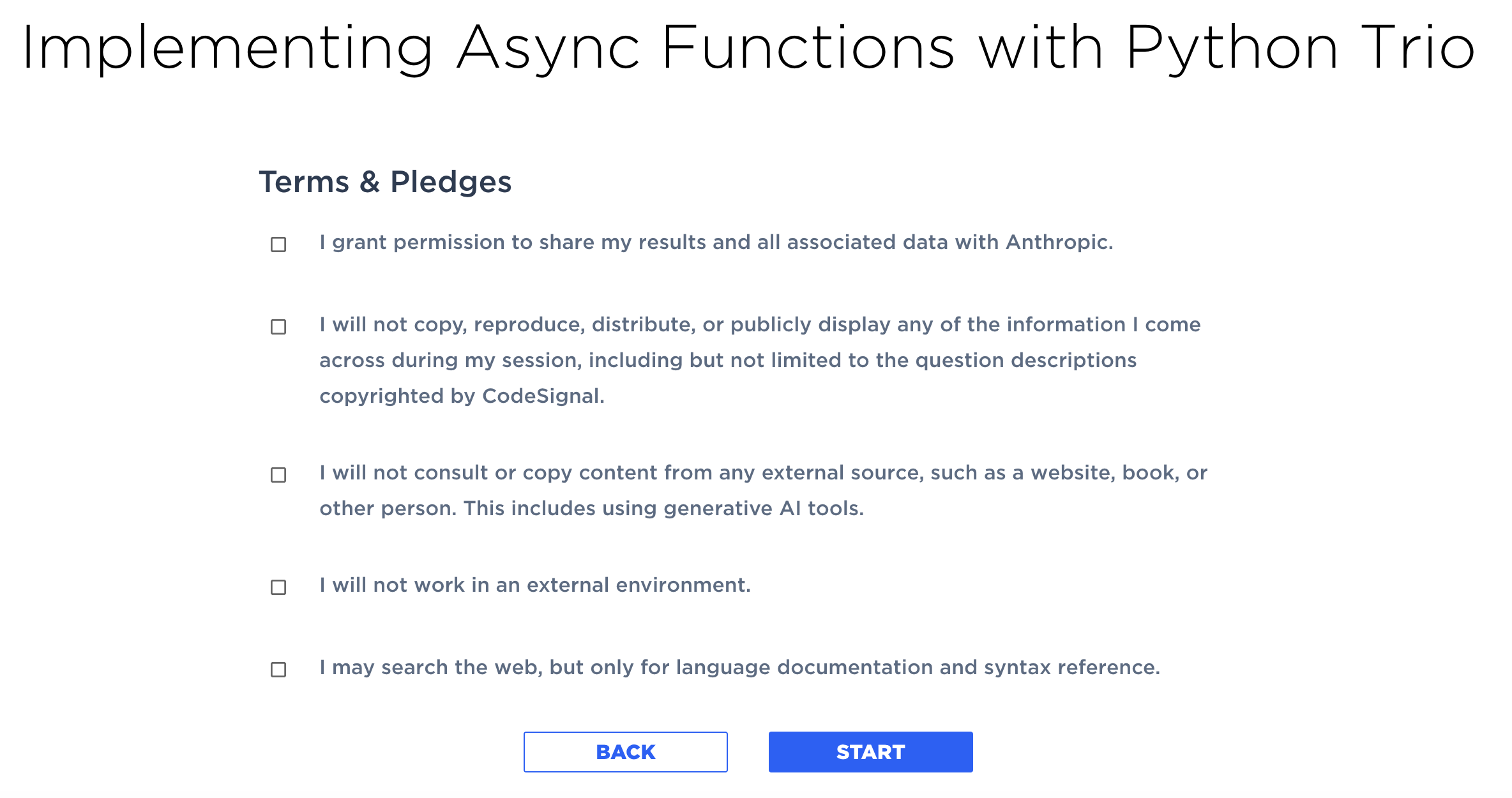}
    \caption{Pledge taken by control group participants: participants agree to not using AI assistance. }
    \label{fig:sc-pledge-control}
  \end{minipage}
  \hfill
  \begin{minipage}{0.45\textwidth}
    \centering
    \includegraphics[width=\textwidth]{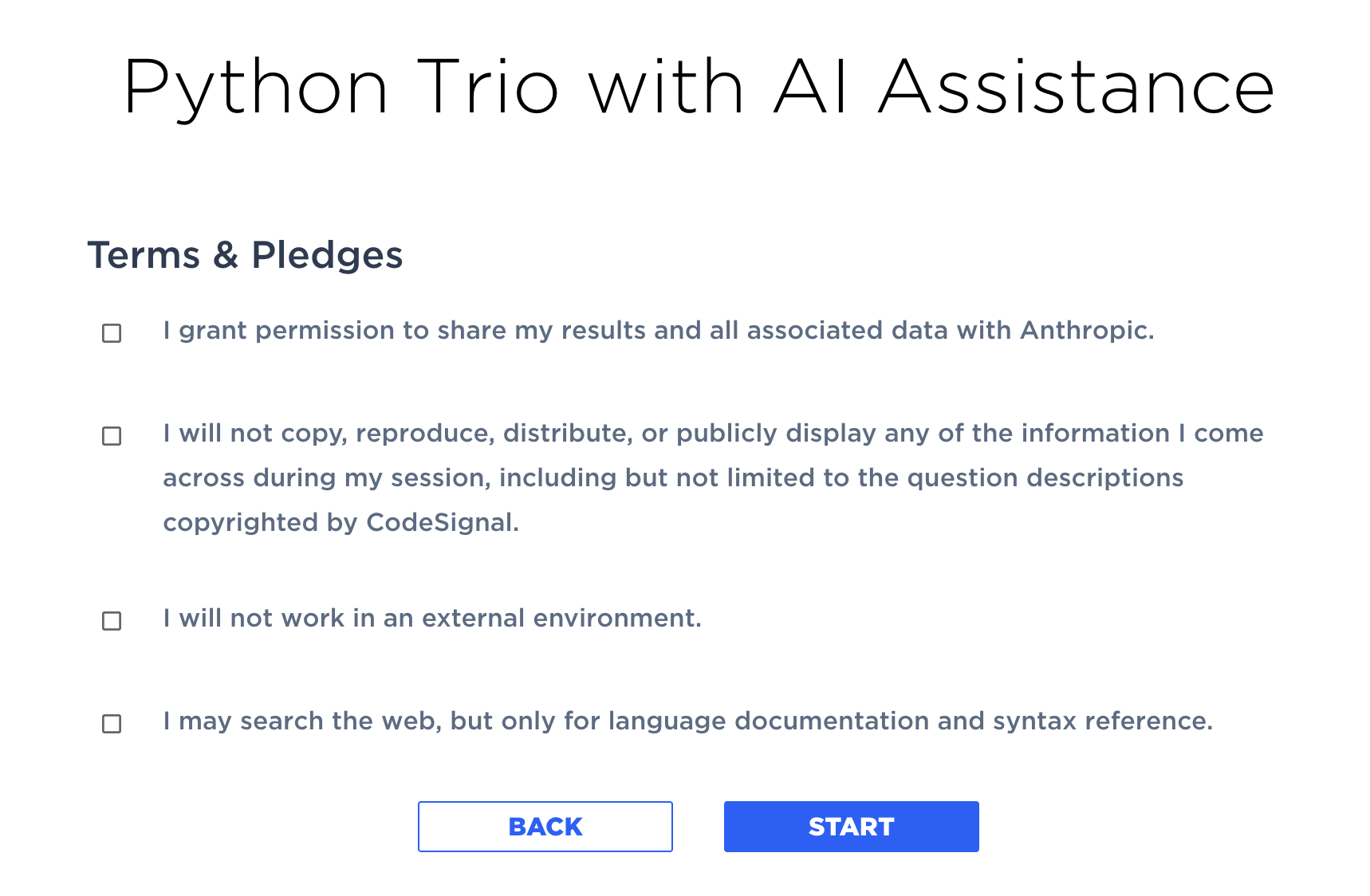}
    \caption{Pledge taken by treatment group participants.}
    \label{fig:sc-pledge-treatment}
  \end{minipage}
\end{figure}

\begin{figure}[h]
  \centering
  \begin{minipage}{0.45\textwidth}
    \centering
    \includegraphics[width=\textwidth]{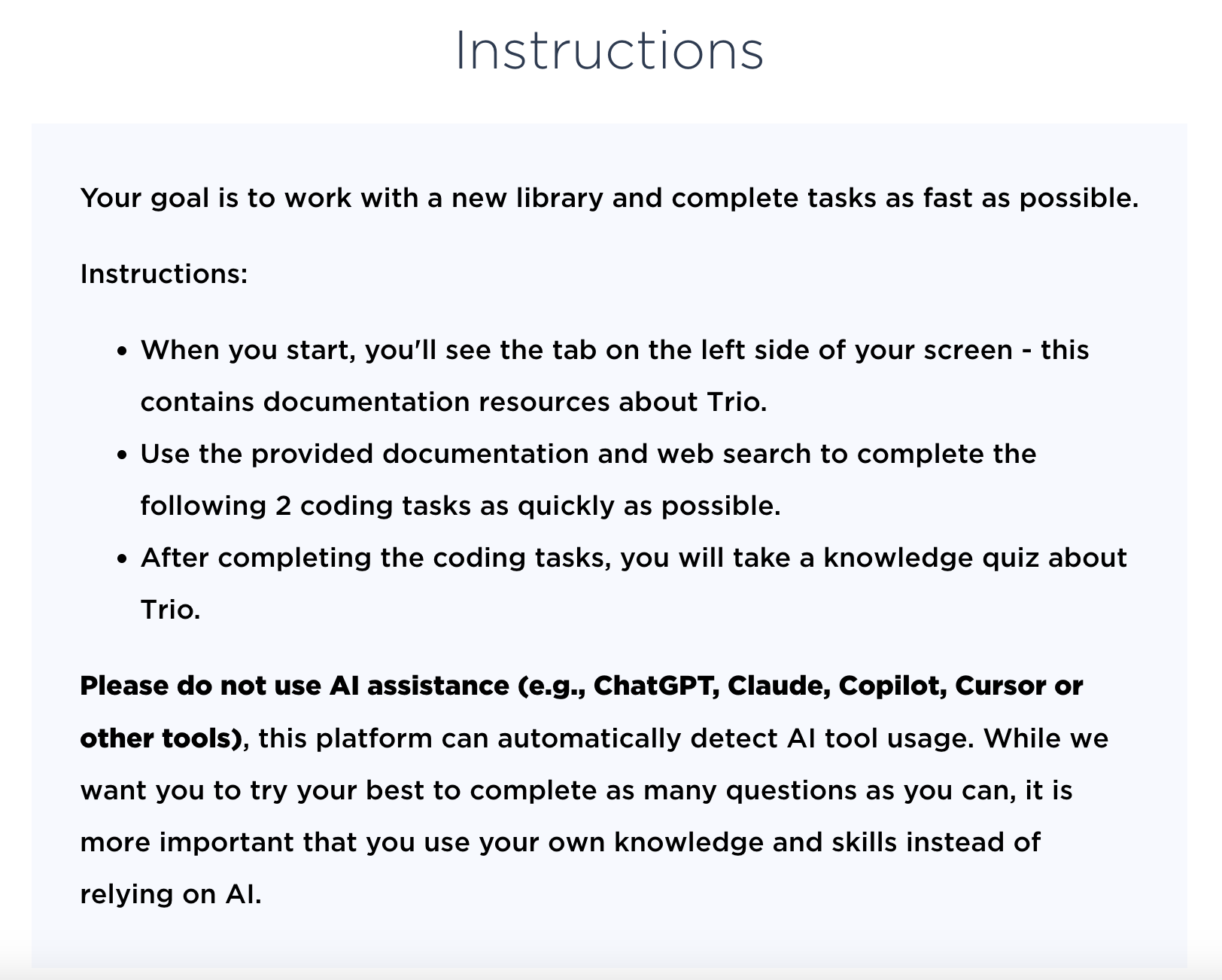}
    \caption{Instructions given to control group participants. We heavily emphasize not using AI tools.}
    \label{fig:sc-instructions-control}
  \end{minipage}
  \hfill
  \begin{minipage}{0.45\textwidth}
    \centering
    \includegraphics[width=\textwidth]{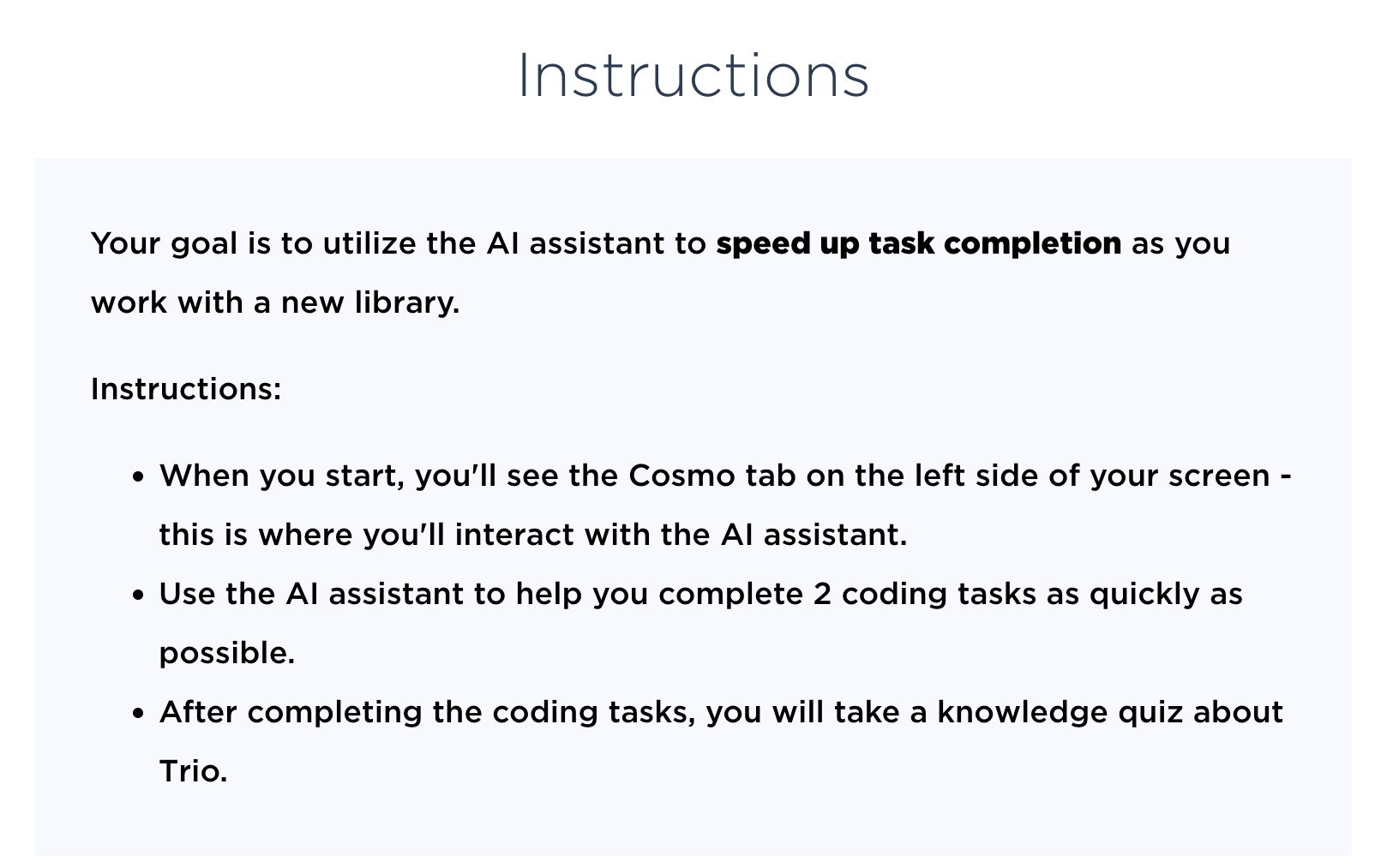}
    \caption{Instructions given to treatment group participants. This group was encouraged to use the AI assistant to complete the task as quickly as possible.}
    \label{fig:sc-instructions-treatment}
  \end{minipage}
\end{figure}

\begin{figure}
    \centering
    \includegraphics[width=\linewidth]{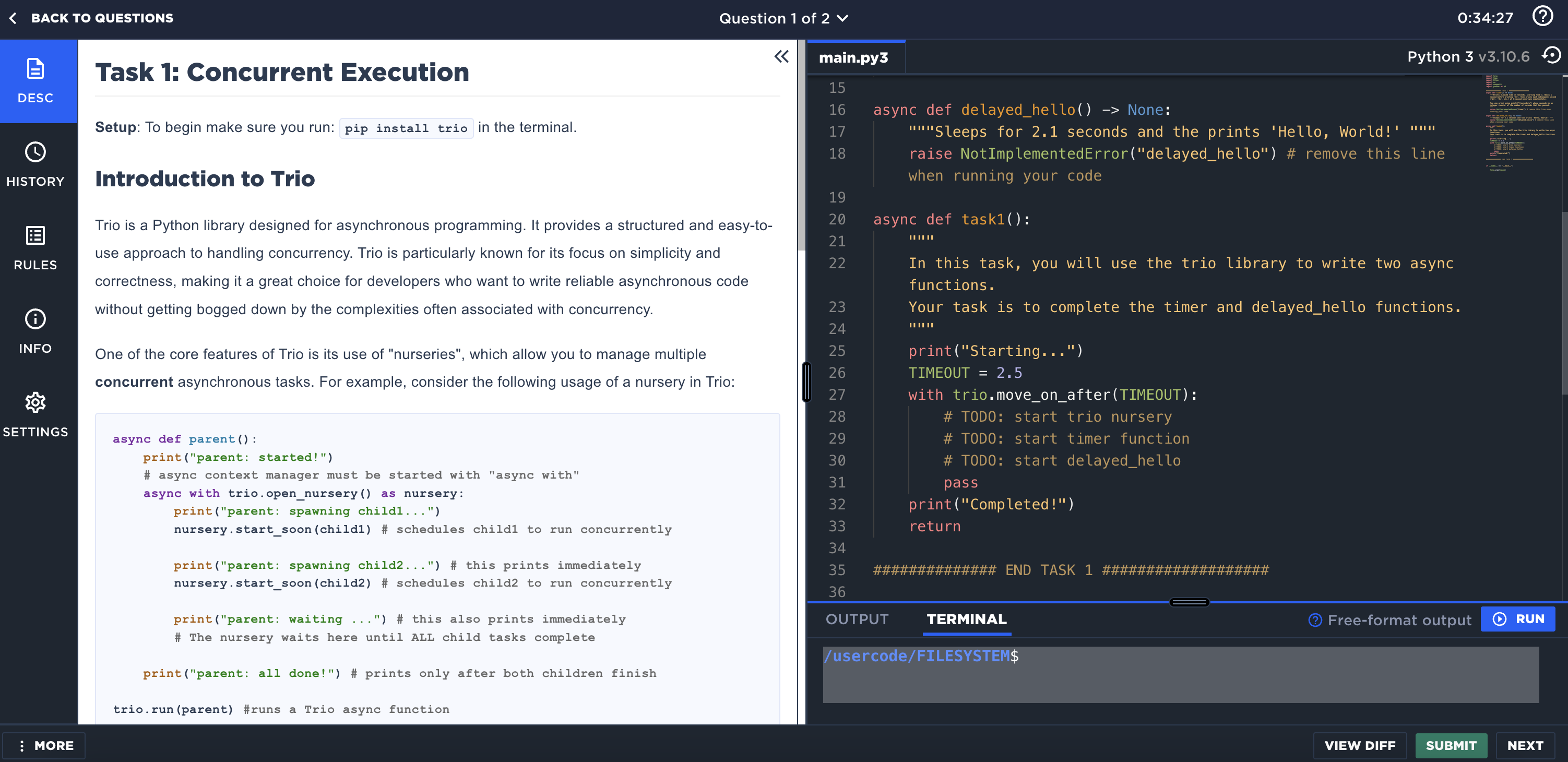}
    \caption{Screenshot of the task platform in the control condition. The instructions are on the left and the coding editor is on the right.}
    \label{fig:sc-interface-control}
\end{figure}

\begin{figure}
    \centering
    \includegraphics[width=\linewidth]{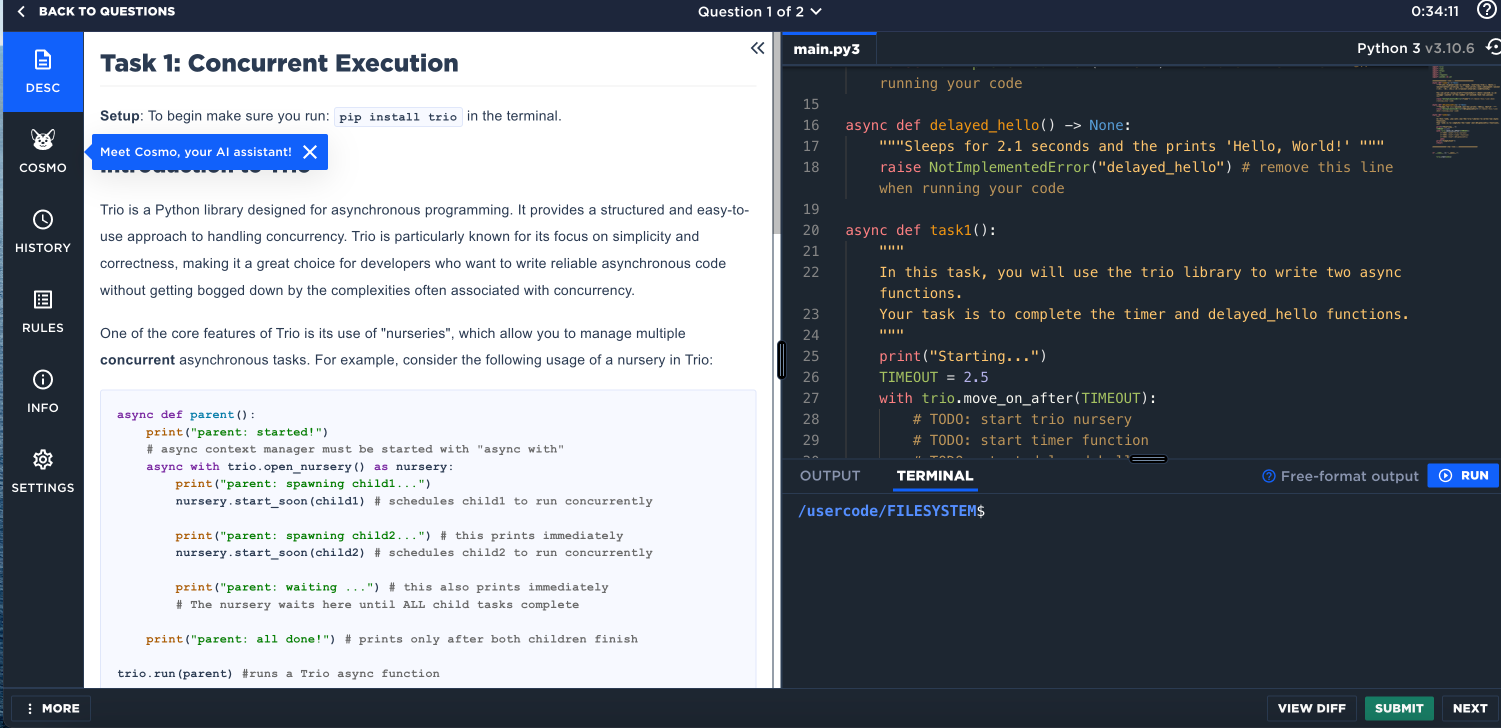}
    \caption{Screenshot of the task platform in the AI condition. The instructions are on the left and the coding editor is on the right. There is a nudge to use the AI assistant on the left tool plane.}
    \label{fig:sc-interface-ai-instruct}
\end{figure}

\begin{figure}
    \centering
    \includegraphics[width=\linewidth]{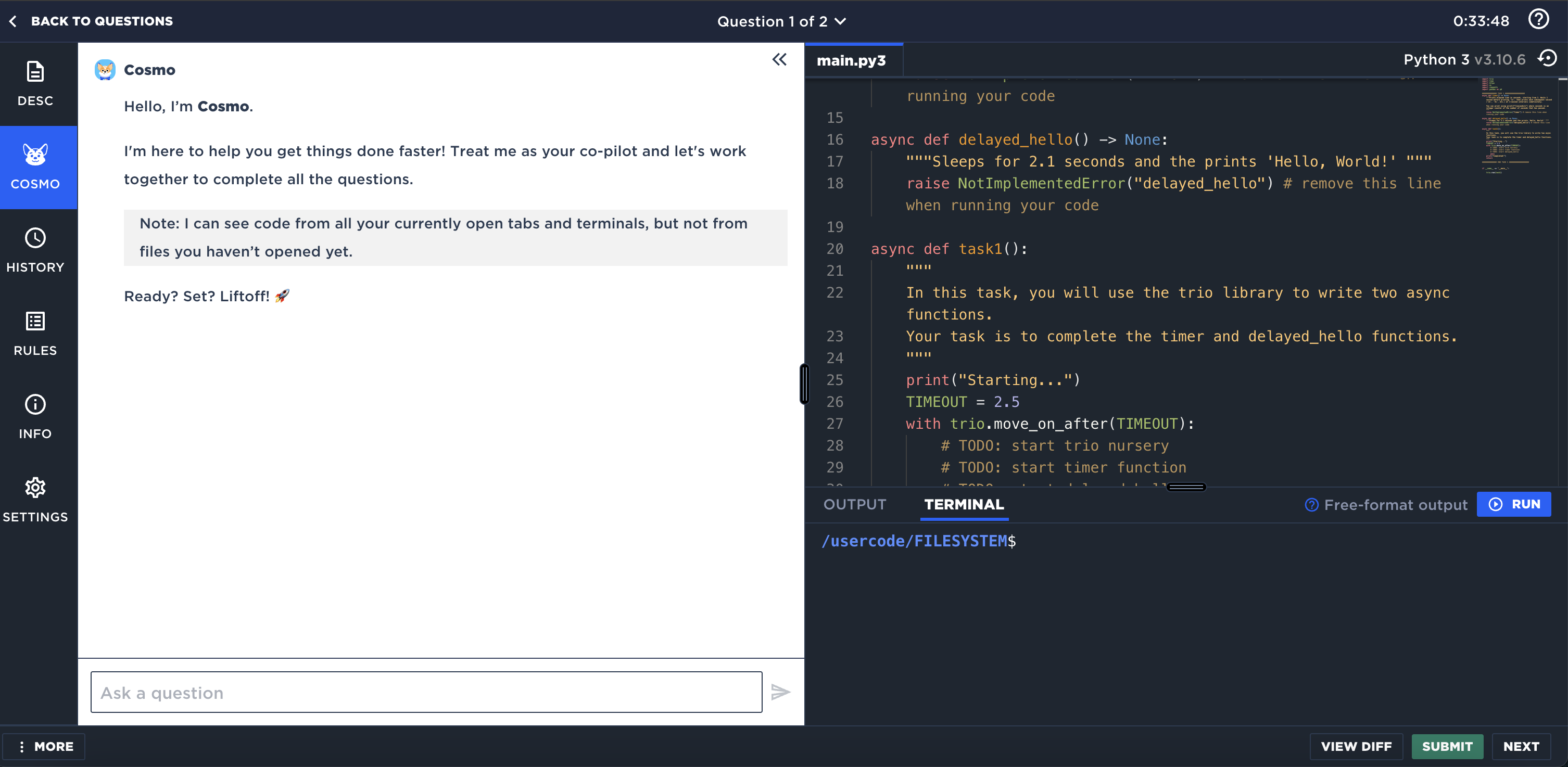}
    \caption{Screenshot of the task platform when interacting with AI Assistant.}
    \label{fig:sc-interface-ai-assistant}
\end{figure}

\end{document}